\documentclass{aa}
\pdfoutput=1
\usepackage{graphicx}
\usepackage{txfonts}
\usepackage{multirow}
\usepackage{natbib}
\usepackage{color}
\usepackage{amsmath}
\newcommand{\masyr}{${\rm mas/yr}$}

\newcommand{\gaia}{{\it Gaia~}}

\begin{document}

   \title{Stellar 3-D kinematics in the Draco dwarf spheroidal galaxy}

   \subtitle{}

   \author{D. Massari\inst{1,2,3}
          \and
          A. Helmi\inst{1}
          \and
          A. Mucciarelli\inst{2,3}
          \and
          L. V. Sales\inst{4}
          \and
          L. Spina\inst{5}
          \and
          E. Tolstoy\inst{1}
          }

   \institute{Kapteyn Astronomical Institute, University of Groningen, NL-9747 AD Groningen, Netherlands\\
              \email{massari@astro.rug.nl}  
         \and
             Dipartimento di Fisica e Astronomia, Universit\`{a} degli Studi di Bologna, Via Gobetti 93/2, I-40129 Bologna, Italy
         \and
             INAF - Osservatorio di Astrofisica e Scienza dello Spazio di Bologna, Via Gobetti 93/3, I-40129 Bologna, Italy
         \and
             Department of Physics and Astronomy, University of California Riverside, 900 University Ave., CA92507, US
         \and
             Monash Centre for Astrophysics, School of Physics and Astronomy, Monash University, VIC 3800, Australia
             }

   \date{Received April 4th 2019; accepted November 12th 2019}


  \abstract
   {}
   {We present the first three-dimensional internal motions for individual stars in the Draco dwarf spheroidal galaxy.}
   {By combining first-epoch {\it Hubble Space Telescope} observations
     and second-epoch \gaia Data Release 2 positions, we measured the
     proper motions of $149$ sources in the direction of Draco. We
     determined the line-of-sight velocities for a sub-sample of $81$
     red giant branch stars using medium resolution spectra acquired
     with the DEIMOS spectrograph at the Keck II telescope. Altogether,
     this resulted in a final sample of $45$ Draco members with
     high-precision and accurate 3D motions, which we present as a table in this paper.}
   {Based on this high-quality dataset, we determined the velocity
     dispersions at a projected distance of $\sim120$~pc from
     the centre of Draco to be $\sigma_{R} =11.0^{+2.1}_{-1.5}$~km/s,
     $\sigma_{T}=9.9^{+2.3}_{-3.1}$~km/s and
     $\sigma_{LOS}=9.0^{+1.1}_{-1.1}$~km/s in the projected radial,
     tangential, and line-of-sight directions. This results in a
     velocity anisotropy $\beta=0.25^{+0.47}_{-1.38}$ at
     $r \gtrsim120$~pc.  Tighter constraints may be obtained using 
the spherical Jeans equations and
     assuming constant anisotropy and Navarro-Frenk-White (NFW) mass profiles, 
     also based on the assumption that the 3D velocity dispersion should be lower than$\approx 1/3$ of the escape velocity of the system. In this case, we constrain the maximum circular velocity
     $V_{max}$ of Draco to be in the range of $10.2-17.0$ km/s. The
     corresponding mass range is in good agreement with previous
     estimates based on line-of-sight velocities only.}
   {Our Jeans modelling supports the case for a cuspy dark matter
     profile in this galaxy. Firmer conclusions may be drawn by
     applying more sophisticated models to this dataset and with new datasets from
     upcoming \gaia releases.}

   \keywords{Galaxies: dwarf – Galaxies: Local Group – Galaxies: kinematics and dynamics – Proper motions –   Techniques: radial velocities}
   
   \maketitle
%

\section{Introduction}

 The success of $\Lambda$-cold dark matter ($\Lambda$CDM) cosmology relies on its ability to describe many of the observed global properties
 of the Universe, from the cosmic microwave background (\citealt{planck14}) to large-scale structure
 (\citealt{springel06}). However, this model is subject to some inconsistencies when considering
 the properties of dark matter haloes on small cosmological scales, such as dwarf galaxies.
 One example is the so-called cusp-core problem according to which the observed 
 internal density profile of dwarf galaxies is less steep than that predicted by CDM simulations (\citealt{flores94, moore94}).
 While several solutions have been proposed to explain the evolution of cusps into cores based on the interaction with baryons 
 (e.g. \citealt{navarro96, read05, mashchenko08}), it remains critical to directly measure
 the dark matter density profile in these small stellar systems.
 
 One of the best ways to do this is to measure the stellar kinematics in dark-matter dominated dwarf spheroidal
 satellites of the Milky Way. 
 Dwarf spheroidals are well-suited for the investigation of the behaviour of dark matter  as the weak stellar feedback
 they have experienced is not likely to have significantly perturbed the shape of their gravitational potential, particularly for objects with $M_\star \lesssim 10^6$~M$_\odot$ \citep[see e.g.][and references therein]{Fitts2017}.
 Thus far, many investigations have tried to exploit line-of-sight (LOS) velocity measurements in these systems
 in combination with Jeans modelling to assess whether cuspy dark matter profiles, such as Navarro-Frenk-White profiles (NFW, \citealt{nfw96}) provide a better fit to the data 
 than cored profiles (e.g. \citealt{burkert95}). However, the results have been conflicting, sometimes favouring the former case 
 (e.g. \citealt{strigari10, jardel13}) and sometimes the latter (e.g. \citealt{gilmore07}), or concluding that both are consistent with
 the observations (e.g. \citealt{battaglia08, breddels13, strigari17}).
 
 Most of these studies are affected by the mass-anisotropy degeneracy (\citealt{binney82}), which prevents an unambiguous
 determination of the dark matter density (\citealt{walker13}) given LOS velocity measurements only.
 However, internal proper motions in distant dwarf spheroidal galaxies are now becoming possible
 (e.g. \citealt{massari18}, hereafter M18) and so we can break this degeneracy. This is thanks to the outstanding astrometric capabilities
 of the {\it Hubble Space Telescope} (HST, see e.g. \citealt{bellini14} and the series of papers from
 the HSTPROMO collaboration) and \gaia (\citealt{brown16, prusti16, brown18}). The combination of these two facilities provides sub-milliarcsec positional precision
 and a large temporal baseline, enabling the measurement of proper motions in distant ($>80$ kpc) stellar systems with
 a precision of $\sim10$ km/s (e.g. \citealt{massari17}, M18).
 
 Despite the measurement of the proper motions in the case of the Sculptor dwarf spheroidal galaxy (M18), 
 the limited number of stars (ten) with measured 3D kinematics resulted in uncertainties too large to
 pin down whether the profile is cusped or cored for that galaxy (\citealt{strigari17}).
 In this paper, we try to obtain more precise 3D kinematics for stars in the Draco dwarf spheroidal
 by measuring proper motions from the combination of HST and \gaia positions,
 and by combining them with LOS velocities purposely obtained from observations recently undertaken with the 
 Deep Imaging Multi-Object Spectrograph (DEIMOS, \citealt{faber03}) at the Keck II telescope.
 
 Draco is an ideal dwarf spheroidal for tackling the cusp-core problem as it most likely maintained a pristine dark matter halo,
 having stopped its star-formation about 10 Gyr ago (\citealt{aparicio01}) and as  one of the most dark matter dominated 
 satellites of the Milky Way (\citealt{kleyna02, lokas04}), with no sign of tidal disturbances (\citealt{segall07}). 
 \cite{read18}  recently presented the results of a dynamical modelling that 
 exploit about $500$ LOS velocity measurements and favour a cuspy dark matter profile. Here we will test this conclusion based
 on 3D kinematics. This is the second galaxy for which this kind of study has been undertaken.

\begin{figure}
 \centering%
\includegraphics[width=\columnwidth]{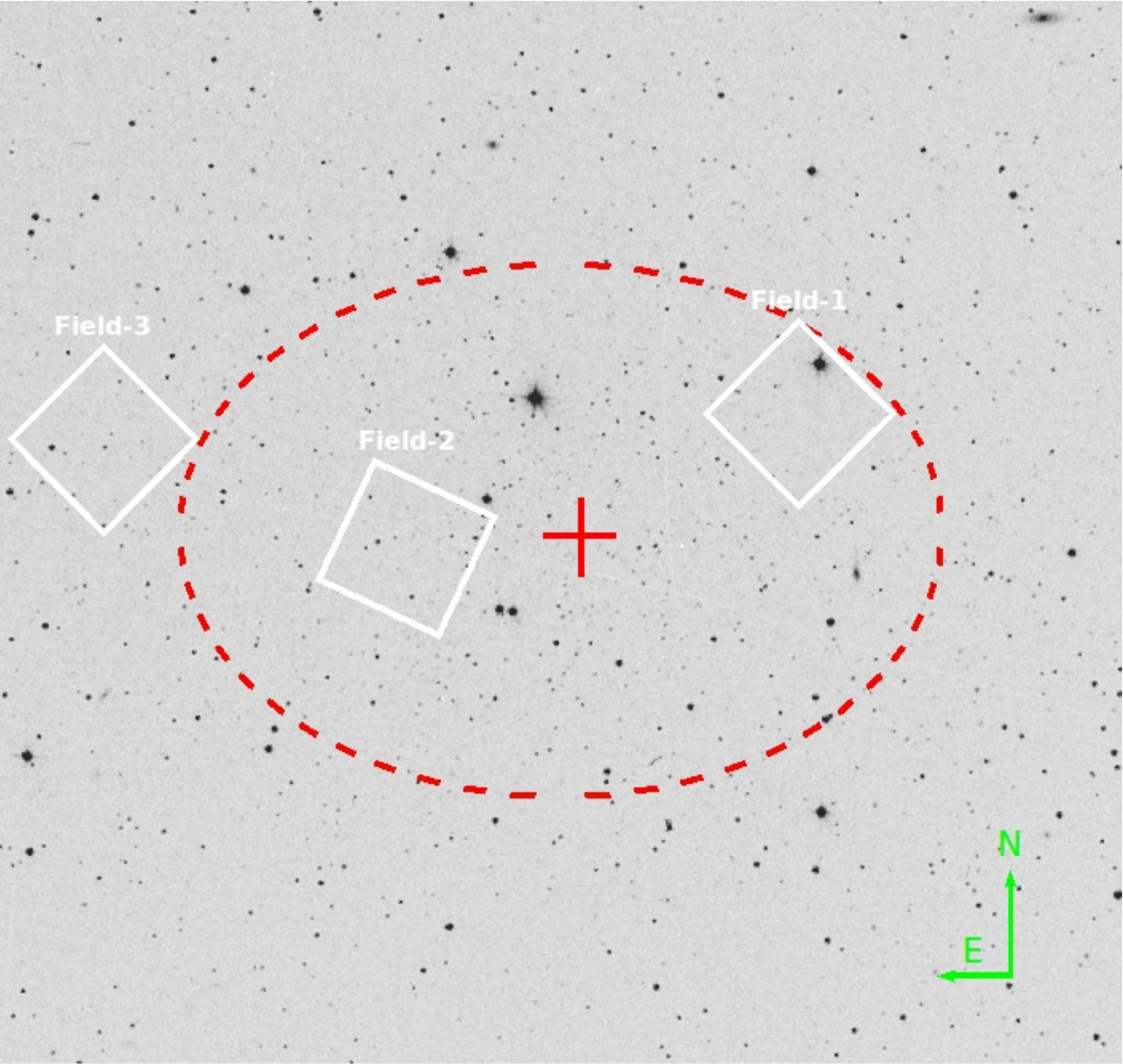}
\caption{Location on the sky of the three HST fields (white boxes). 
Draco's centre, half-light radius (shown as red cross and dashed red circle, respectively) and ellipticity are taken from \cite{munoz18}.
The background $\sim15\arcmin \times 15\arcmin$ image comes from the Sloan Digital Sky Survey.}\label{fields}
\end{figure}

\begin{figure}
 \centering%
\includegraphics[width=\columnwidth]{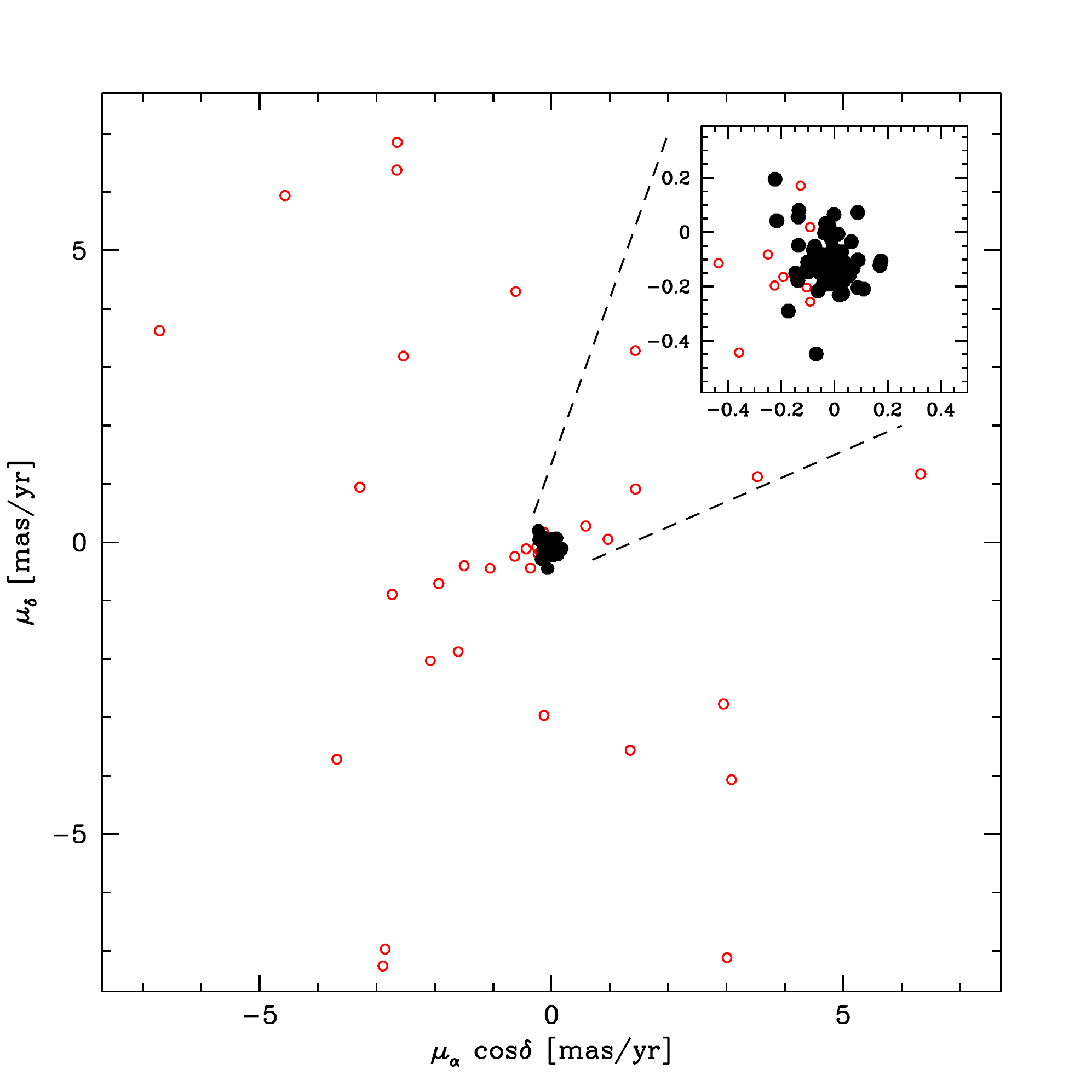}
\caption{Vector point diagram for the sources in the direction of Draco with measured proper motions. Likely members are shown as black symbols,
while likely foreground contaminants are marked with red empty symbols. The region zoomed around the bulk of
likely Draco members is highlighted in the inset, where the few red symbols represent stars fainter than $G=20.8$ which are excluded from the analysis
(see Sect.\ref{pms}).}\label{vpd_wide}
\end{figure}

\section{Proper motions}\label{pms}

Relative proper motions for Draco stars are measured from the combination of HST and \gaia positions.
The first-epoch positional measurements come from observations acquired with the Wide Field
Channel (WFC) of the Advanced Camera for Survey (ACS) on board
HST. This camera consists of two $2048 \times 4096$ pixel detectors, with a 
pixel scale of $\sim0.05\arcsec\,$pixel$^{-1}$, and separated by a gap of about 50 pixels 
for a total field of view (FoV) $\sim200\arcsec \times 200\arcsec$.
The data set (GO-10229, PI: Piatek) consists of three pointings, located at $\sim 4\arcmin$,
$\sim 5\arcmin$ and $\sim 11.7\arcmin$ distances from the dwarf nominal centre (\citealt{munoz18}),
as shown in Fig.~\ref{fields}. 
Each pointing was observed $19$ times, with each exposure having a duration of $430$ s. 
Exposures in Field-1 were taken in the F555W filter, while those in the other two fields were acquired
in the F606W filter. The observations took place on the 30th and 31st of October, 2004.

The data reduction was performed using the \texttt{img2xym$\_$WFC.09$\times$10} programme from \cite{jayking06}
on \_FLC images, which were corrected
for charge transfer efficiency (CTE) losses by the pre-reduction
pipeline adopting the pixel-based correction described in \cite{jaybedin10,ubedajay}. 
Each chip of each exposure was analysed independently using a pre-determined, filter-dependent model of the Point Spread
Function (PSF), and for each chip we obtained a catalogue of all the unsaturated sources with positions, instrumental magnitudes, and 
PSF fitting-quality parameter. 
We used filter-dependent geometric distortions (\citealt{jayacs}) to correct the stellar positions.
The 19 catalogues of each field and chip were then rotated to be aligned with the equatorial
reference frame and cross-matched using the stars in common for at least 15 of them. 
Once the coordinate transformations were determined,
a catalogue containing average positions, magnitudes, and corresponding uncertainties (defined as the rms of the
residuals around the mean value) for all of the sources detected in at least four individual exposures
was created.
In this way, at the end of the reduction we have six ACS/WFC catalogues, one per chip and field, 
ready to be matched with the \gaia second-epoch data.

Second-epoch positions are provided by the \gaia second data release (DR2, \citealt{brown18}).
This is a significant upgrade with respect to our previous work on the Sculptor dwarf spheroidal
(M18) as DR2 is more complete than DR1 and allows us to have more faint stars
in common between the two epochs because DR2 positions are more accurate and precise (the \gaia astrometric
solution improves as more observations are collected with time, see \citealt{lindegren18, arenou18}).
From the \gaia archive\footnote a , we retrieved a catalogue with J2015.5
positions, related uncertainty and correlations, as well as magnitudes and astrometric excess noise for all the
sources within a distance of $20\arcmin$ from the centre of Draco. In order to exclude 
sources with a clearly problematic solution from the analysis, we discarded all those with positional errors larger than $2$ mas
(whereas the median positional uncertainty is $0.4$ mas).
This dataset provides a total temporal baseline for proper-motion measurements of $10.593$ years.

As the final step in measuring the stellar proper motions, we transform HST first-epoch positions
to the \gaia reference frame using a six-parameter linear transformation as described in M18.
The difference between \gaia- and HST-transformed positions divided by the temporal baseline
provides our proper-motion measurements, whereas the sum in quadrature between the two epochs'
positional errors divided by the same baseline provides the corresponding proper-motion uncertainties.
We refined the coordinate transformations iteratively, each time using only likely members
of Draco based on their location in the colour magnitude diagram (CMD) and the previous proper-motion determination.
After three iterative steps, the procedure was converged (no stars were added or lost in the list used to compute
the transformations in subsequent steps) and the final list of stellar proper motions was built, including $149$ sources. We bring these relative proper motions to an absolute reference frame using
the Draco mean absolute motion of ($\mu_{\alpha}\cos(\delta), \mu_{\delta}$)$=(-0.019, -0.145)$ \masyr~ 
as reported in \cite{helmi18}.

\begin{figure}
 \centering%
\includegraphics[width=\columnwidth]{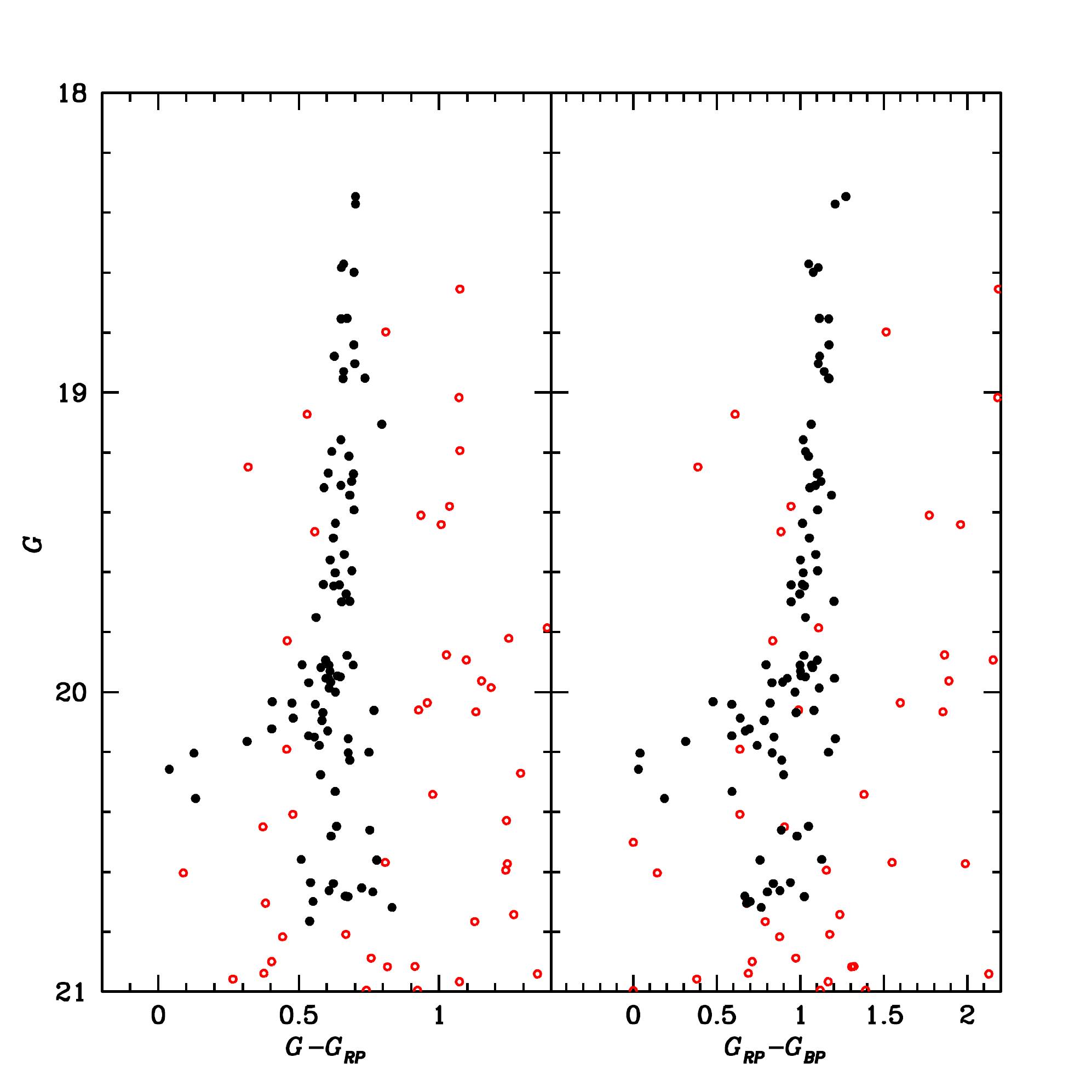}
\caption{\gaia ($G$, $G$-$G_{\rm RP}$) and ($G$, $G_{\rm BP}$-$G_{\rm RP}$) colour magnitude diagrams. Likely members roughly selected from the VPD as 
described in the text are marked with black-filled symbols and correspond to those shown in Fig.~\ref{vpd_wide}.}\label{cmd_wide}
\end{figure}

The proper-motion measurements for all of the sources are shown in the Vector Point Diagram (VPD) in Fig.~\ref{vpd_wide}. The clump is centred
around the Draco mean absolute motion, thus describing the likely members, and clearly separates them from the distribution of likely
foreground stars, which have much larger proper motions. As a consistency check, Fig.~\ref{cmd_wide} shows the 
\gaia ($G$, $G_{\rm RP}$) and ($G$, $G_{\rm BP}$-$G_{\rm RP}$) CMDs for the same sources. Black symbols indicate likely members, roughly selected
as all the stars located within a 1 \masyr~distance from Draco's absolute motion. They describe the well-defined sequences expected for the red giant and horizontal branches of Draco, whereas red symbols are mostly distributed
in regions of the CMD populated by field stars. We note that this selection is not at all refined but has only been carried out
as a first check that the proper-motion measurement procedure worked correctly.
The final kinematic membership will be assessed after coupling the proper motions with radial velocity measurements.\\

 \begin{figure}
 \centering%
\includegraphics[width=\columnwidth]{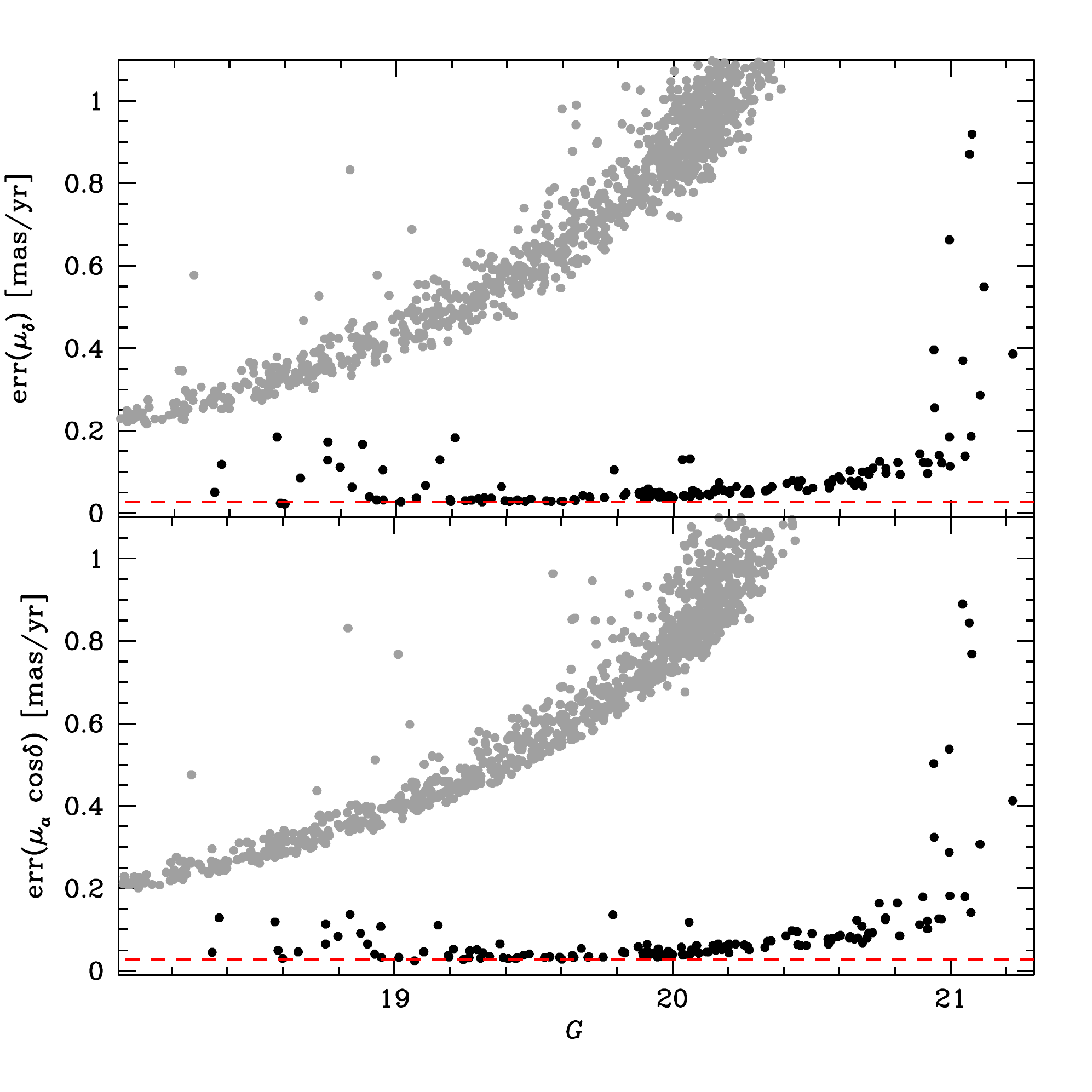}
\caption{Intrinsic uncertainties as a function of \gaia $G$-band magnitude for the proper-motion component along $\alpha$ (lower panel) and $\delta$ (upper panel).
Black-filled symbols are our HST+\gaia estimates, while grey-filled symbols are \gaia DR2 proper-motion uncertainties for sources
in the direction of Draco. The red dashed line corresponds to a velocity of 10 km/s at the distance of Draco, roughly mimicking
the system velocity dispersion.}\label{err}
\end{figure}

Since our goal is to determine the velocity dispersion for the two proper-motion components in the plane of the sky,
it is of primary importance to have all of the uncertainties, both statistical and systematic, under control.

Fig.~\ref{err} shows the distribution of the proper-motion statistical errors as a function of \gaia $G$-band magnitude.
For comparison, grey symbols show \gaia DR2 proper-motion errors (on a baseline of $22$ months) for stars in the direction of the Draco dwarf spheroidal.
The gain in precision obtained through our method is remarkable thanks to the larger baseline of $\sim 126$ month. 
At $G\simeq19.5$, our measurements
are one order of magnitude better than the \gaia DR2 proper motions alone and the improvement is even greater at fainter
magnitudes.

\begin{figure}
\includegraphics[width=\columnwidth]{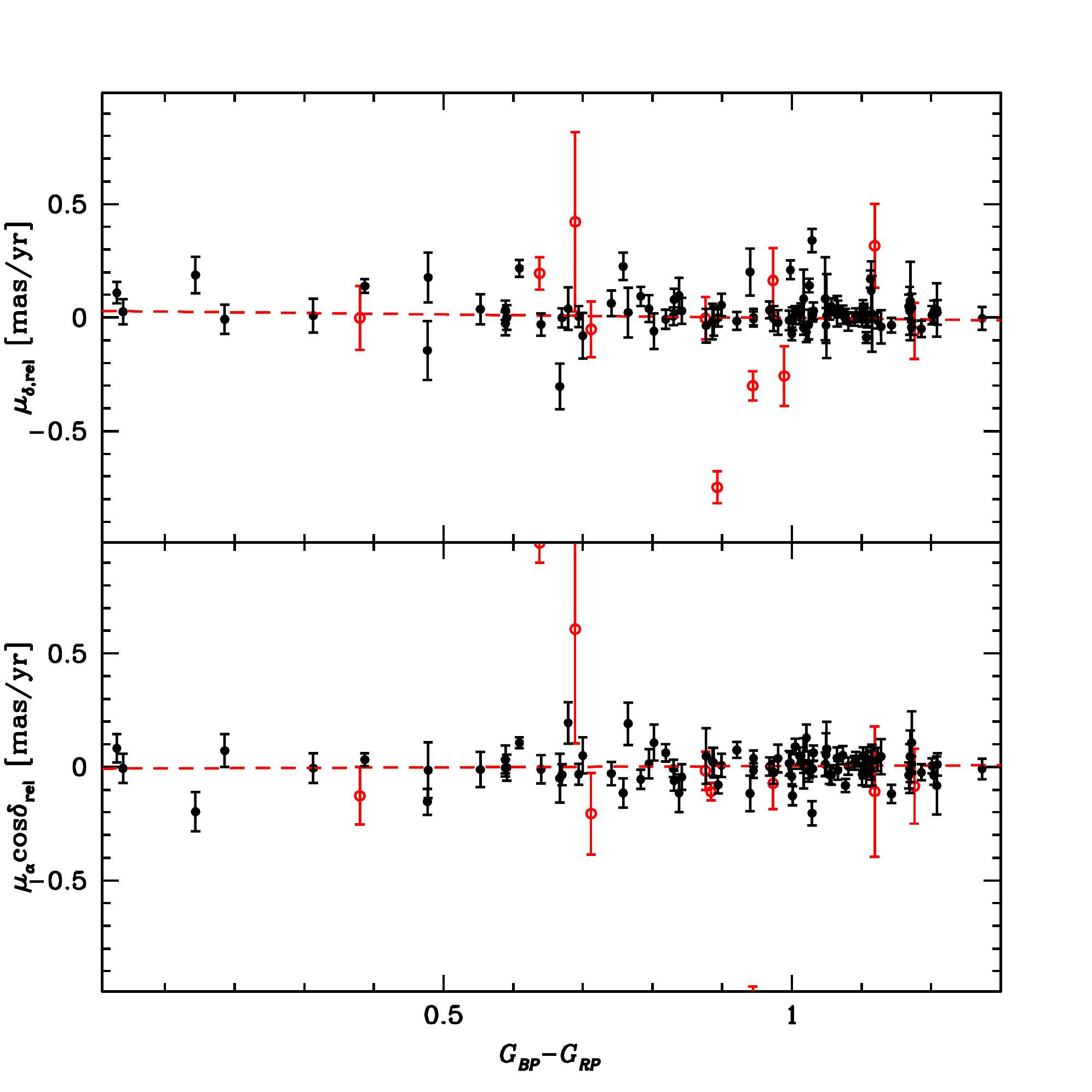}
\caption{\small Trend of the two proper-motion components as a function of \gaia $G_{\rm BP}$-$G_{\rm RP}$ colour (same colour-coding as in Fig.~\ref{cmd_wide}).
The best linear fits (red dashed lines) are not consistent with any trend among the uncertainties.}\label{syscol}
\end{figure}

\begin{figure}
\includegraphics[width=\columnwidth]{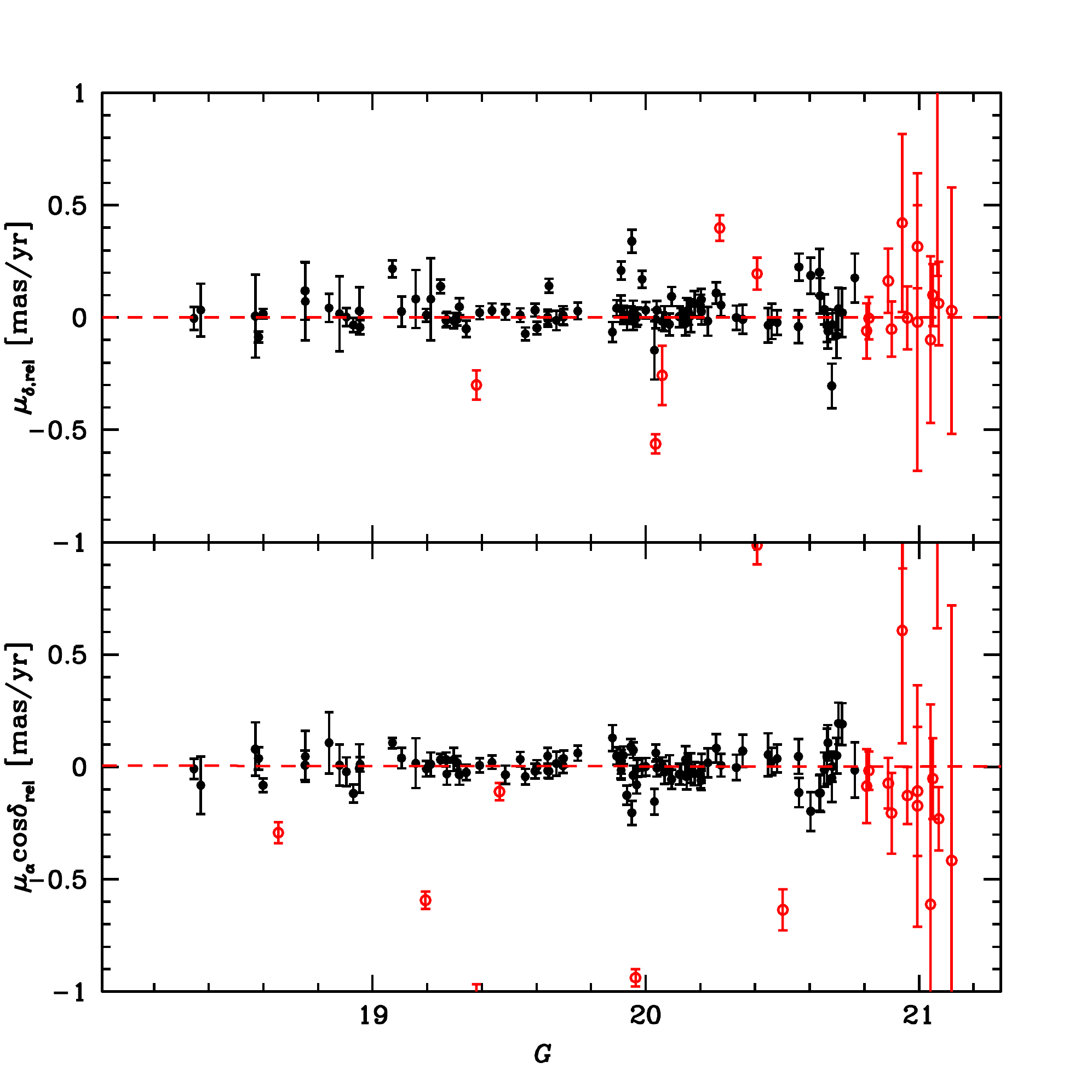}
\caption{\small Trend for two components of the measured proper motions as a function of \gaia $G$-band magnitude (same colour-coding as in Fig.~\ref{cmd_wide}). 
The only systematic effect is visible for stars fainter than $G=20.8$, which are, therefore, excluded from the analysis. Red dashed lines
show the best linear fit after their exclusion and show no trend among the uncertainties.}\label{sys}
\end{figure}

To check for systematic effects, we look for possible trends between our relative proper-motion measurements
and all the parameters entered into the analysis, such as \gaia colours, positions on the sky, HST magnitudes, location on the HST detector, etc.  
We always found consistency with no apparent trends within a 1-$\sigma$ uncertainty. In Fig.~\ref{syscol}, we provide as an example the behaviour of the 
proper motions with respect to the \gaia $G_{\rm BP}$-$G_{\rm RP}$ colour. The only case where we find a significant systematic effect 
is when plotting the relative proper motion as a function of \gaia $G$ magnitude and this is shown in Fig.~\ref{sys}.
This effect is only apparent for stars fainter than $G=20.8$ in the $\mu_{\alpha}\cos(\delta)$ versus {\it G} panel, 
with the proper motions being systematically negative. For this reason, we exclude all of the stars fainter than
this (empty symbols in Fig.~\ref{sys}) from the analysis that follows. Note that these stars are also the ones with the largest 
statistical errors in Fig.~\ref{err} and that because of their faint magnitude, they will also lack a LOS velocity measurement, 
implying that in any case they would not be considered for the dynamical analysis presented later in this paper.

\section{Line-of-sight velocities}\label{rvs}

Line-of-sight velocities (v$_{LOS}$) provide the third kinematic dimension needed for dynamical modelling.
As a first step, we searched the literature for stars with available v$_{LOS}$ measurements included in our proper-motion catalogue.
By using the publicly available samples of \cite{arm95, kleyna02, walker15}, consisting of more than two thousand
measurements, we found a match for only 30 out of 149 stars, all based on the compilation from \cite{walker15}.
This is not surprising, due to the HST field of view being much smaller than the typical 
area sampled by spectroscopic surveys. Moreover, the stars for which we have proper motions are quite faint for spectroscopy
and they can only be observed with 8-m class telescopes instrumentation to achieve a good signal-to-noise ratio (S/N$>10-15$)
within a reasonable length of observing time.

It is for this reason that we targeted all of the stars in our proper-motion sample that are brighter than $G=20.8$ 
in a campaign with the DEIMOS spectrograph at the Keck II telescope.
The strategy adopted to acquire the spectra is the same as that described in \cite{massari14a, massari14b}.
We used the $1200$ line/mm grating, centred at $8000$~\AA~ (to retain the CaII triplet
lines and avoid the inter-CCD gap) and coupled it with the 
GG$495$ and the GG$550$ order blocking filters. The slit width was chosen to be $0.75\arcsec$.
In this way, we covered the range $6500-9500$\AA~ with a spectral resolution of R$\simeq7000$
to sample prominent features like Ca triplet, H$\alpha$ and several strong metallic lines, which are all
well-suited for v$_{LOS}$ determination. 
 
\begin{figure}
\includegraphics[width=\columnwidth]{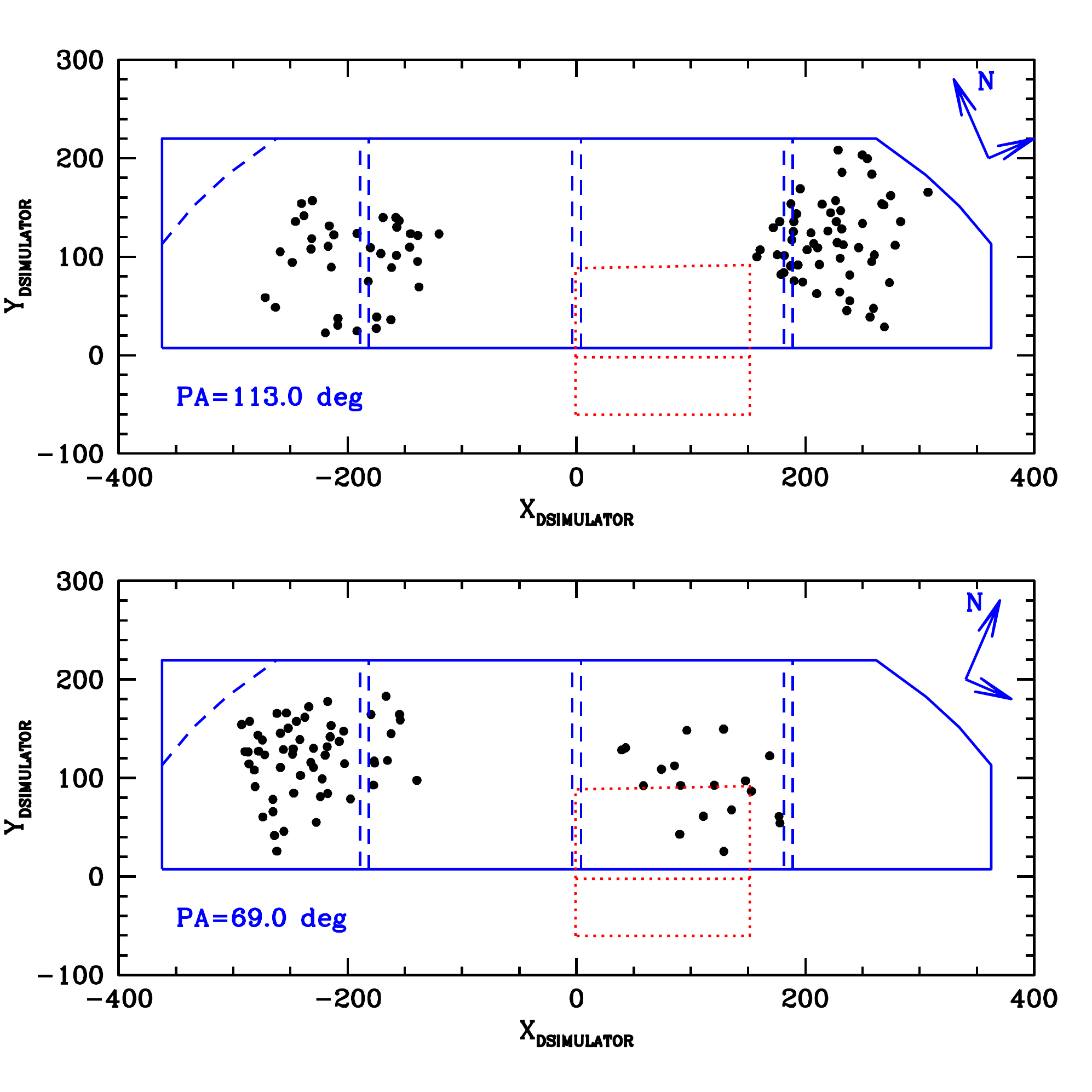}
\caption{\small Here we show two of the DEIMOS masks. Blue lines outline the field of view of DEIMOS,
with dashed lines separating individual chips. Blue arrows show the mask orientation on the sky. The red dotted
square indicates DEIMOS guider. Black-filled symbols mark the available targets.
}\label{masks}
\end{figure}

Since the field of view of each DEIMOS mask is approximately $16\arcmin\times5\arcmin$, two out of three HST fields
can be covered by a single pointing.
Because of the high stellar density in the HST fields and the minimum slit length of 5 arcsec, 
we needed four masks to cover most of the desired targets, and to avoid issues of 
contamination from neighbouring targets. Some of the (faintest) targets had to be dropped because 
they could not fit any allowed mask configuration. Two mask configurations are shown
in Fig.~\ref{masks}.

The observations were performed over two nights, on the 11th and 12th of August 2018 (Project code: U108, PI: Sales).
In order to obtain a v$_{LOS}$ precision comparable to or better than that of the tangential velocities,
we observed each target with at least two $2100$ s long exposures (see Table~\ref{spec_obs}). 

\begin{table}
\caption{DEIMOS spectroscopic observations}             
\label{spec_obs}      
\centering                          
\begin{tabular}{c c c c}        
\hline\hline                 
Mask & Night & t$_{exp}$ (s) & Seeing (\arcsec) \\    
\hline                        
    \multirow{1}{*}{Mask-1A} & August 11th & $2\times2100$ & 0.8 \\
\hline       
    \multirow{3}{*}{Mask-2A} & August 11th & $1\times2100$ & 0.7 \\
     & August 12th & $1\times2400$ & 0.7\\
     & August 12th & $1\times900$ & 0.7\\ 
\hline
    \multirow{1}{*}{Mask-1B} & August 12th & $2\times2400$ & 0.7 \\
\hline
    \multirow{1}{*}{Mask-2B} & August 12th & $2\times2100$ & 0.7 \\
\hline     
\end{tabular}
\end{table}

One-dimensional spectra were extracted using the DEEP2 DEIMOS pipeline (\citealt{cooper12, newman13}).
The software takes calibration (flat fields and arcs) and science images as input to produce output
1D spectra that are wavelength calibrated and combined.
The frames are combined using an inverse-variance weighted algorithm to properly take account of 
different exposure times.

The LOS velocities of the observed stars were measured 
by cross-correlation against a synthetic template spectrum 
using the IRAF task {\tt fxcor}. 
The template spectrum has been calculated with the 
SYNTHE code (\citealt{sbordone04,kurucz05}), adopting atmospheric 
parameters typical for a metal-poor giant star in Draco, namely 
T$_{\rm eff}\simeq4500$ K, log~g$=1.5$, [Fe/H]$=-2.0$. 
The synthetic template is convoluted with a Gaussian profile 
corresponding to the instrumental resolution of DEIMOS.
Spectra that were either too contaminated by neighbours, or with too low of a S/N
to be used to obtain a v$_{LOS}$ measurement, were discarded from the analysis.
All the v$_{LOS}$ have been corrected for heliocentric motion. 
To check for possible wavelength calibration systematics,
we also cross-correlated the observed spectra with synthetic spectra for the Earth's atmosphere calculated with the code TAPAS (\citealt{bertaux14})
around the atmospheric absorption Fraunhofer A band ($7600$-$7700$~\AA).
We found an average offset of $2.8\pm0.2$ km/s, which was applied to each spectrum.

\begin{figure}
\includegraphics[width=\columnwidth]{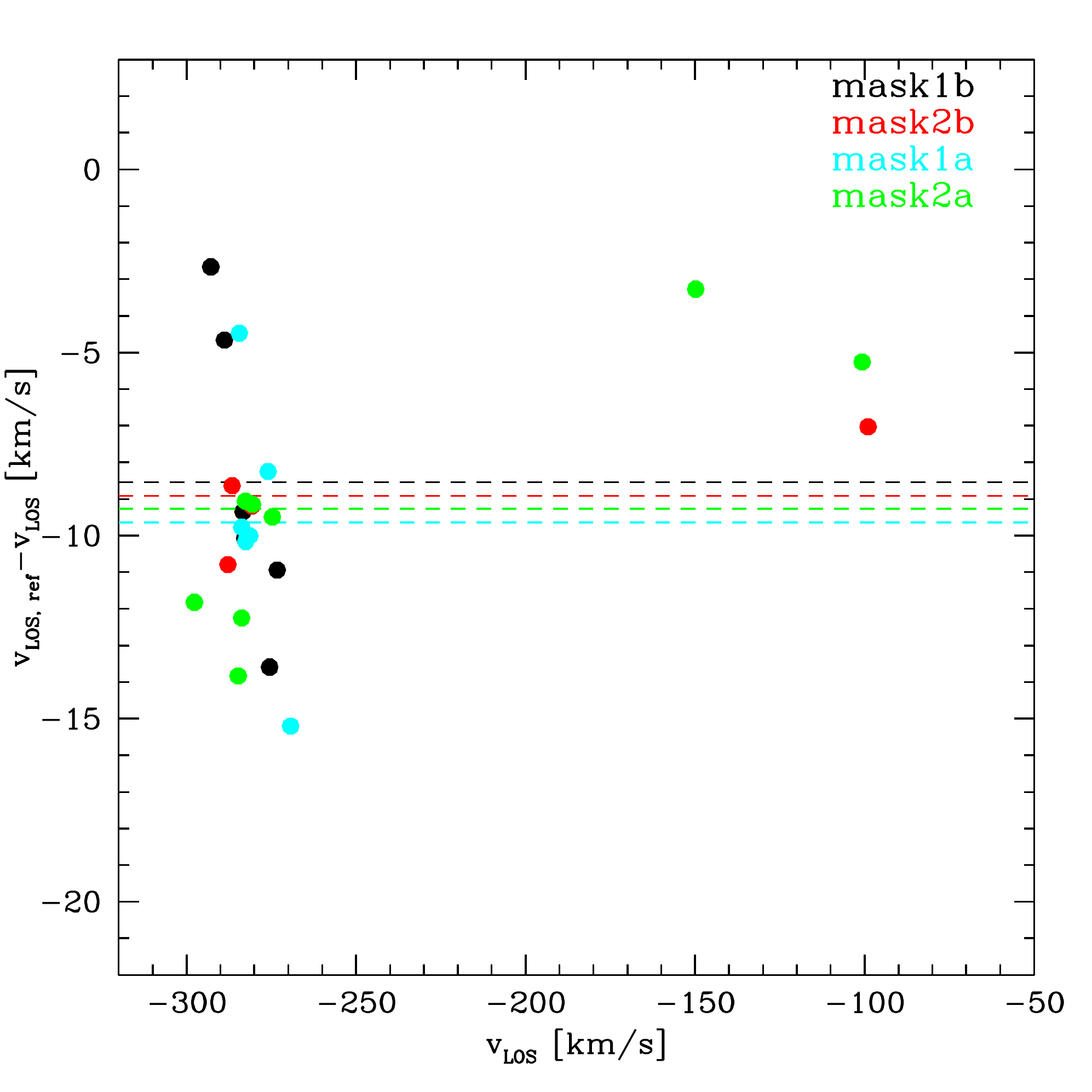}
\caption{\small Offset between our measurements and those of \cite{walker15} for the targets in common.
The colours mark different masks as indicated by the labels. Dashed lines are the adopted zero-points.}\label{zp}
\end{figure}

Because of our desire to combine our set of v$_{LOS}$ with that of \cite{walker15},
a number of slits in each mask were used to observe targets in common to
assess systematic offsets between the two samples.
We thus calibrated the velocities
obtained for each DEIMOS mask to the measurements by \cite{walker15}. The offsets (ZP)
are shown in Fig.~\ref{zp}, amounting to
$ZP_{1A}=-9.6$ km/s ($\sigma_{ZP,1A}=3.5$ km/s); $ZP_{1B}=-8.5$ km/s ($\sigma_{ZP,1B}=4.1$ km/s); 
$ZP_{2A}=-9.3$ km/s ($\sigma_{ZP,2A}=3.6$ km/s); $ZP_{2B}=-8.9$ km/s ($\sigma_{ZP,2B}=1.6$ km/s).
They are all consistent within 1$\sigma$, yet we decided to apply to the targets of each mask the appropriate zero-point (ZP) correction.
We also checked that by applying the same zero-point to all of the target v$_{LOS}$, the results of the study
do not change as the net effect is a change in the v$_{LOS}$ dispersion by $\sim0.05$ km/s.

\begin{figure}
\includegraphics[width=\columnwidth]{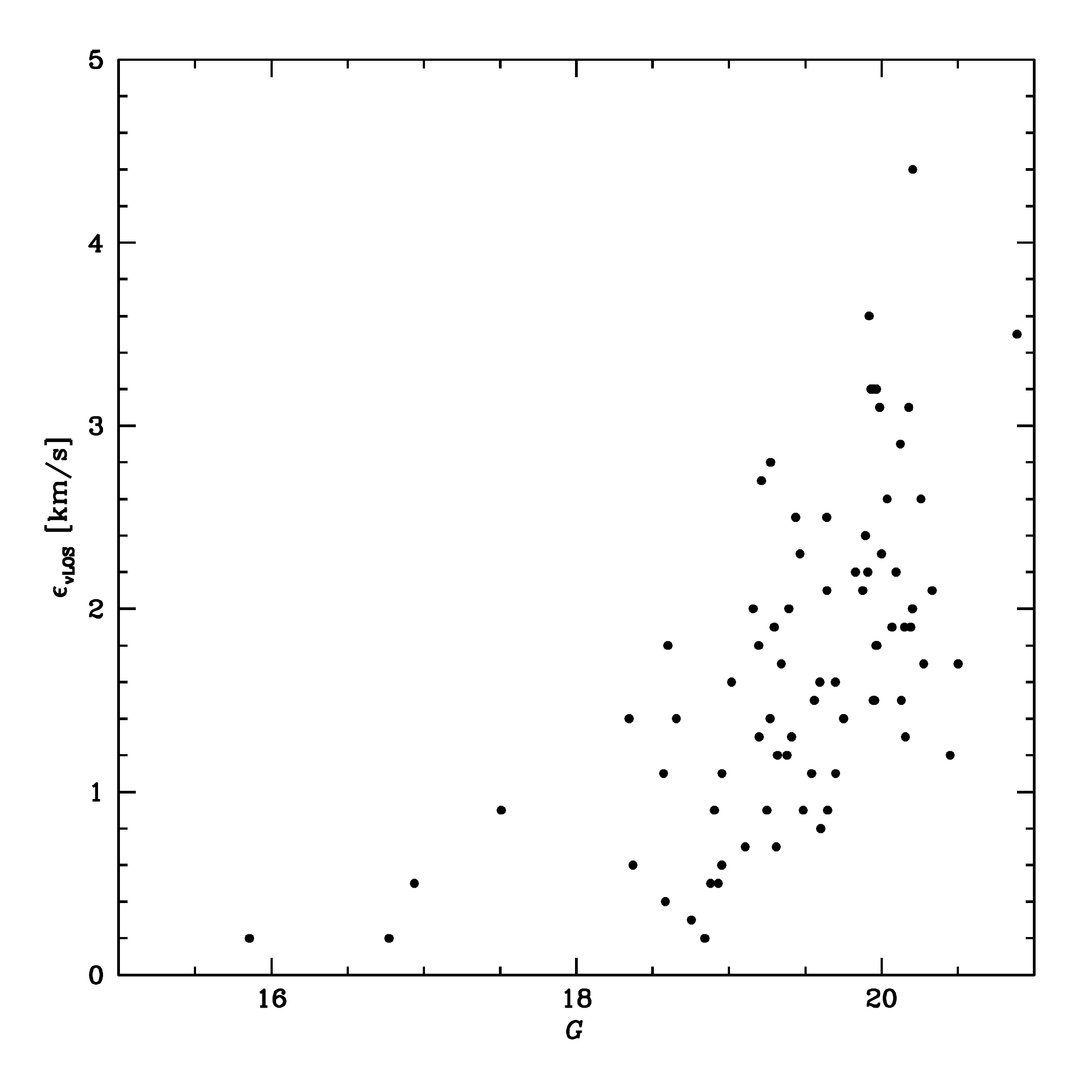}
\caption{\small Intrinsic uncertainties on the LOS velocity as function of \gaia G-band magnitude.}\label{rv_err}
\end{figure}

The intrinsic uncertainties in the measured v$_{LOS}$ have been estimated with Monte Carlo simulations following the approach described in \citet{simongeha07}.
We added Poisson noise to each pixel of one of the extracted 1D spectra
with the highest S/N in order to reproduce different noise conditions.
This procedure has been repeated to cover the entire range of
S/N measured in the observed spectra, in steps of $\Delta$S/N$=10$.
For each set of 300 "noisy"\ spectra with a given S/N, v$_{LOS}$ was measured
with the procedure described above and the associated uncertainty $\epsilon_{LOS}$ was
computed as the dispersion around the mean v$_{LOS}$ value. In this way, we can 
derive a S/N-$\epsilon_{LOS}$ relation and use it to provide a v$_{LOS}$ uncertainty to each observed spectrum, 
based on its S/N. The distribution of the intrinsic errors as a function of \gaia {\it G} magnitude is given
in Fig.~\ref{rv_err}.

As shown in \cite{simongeha07} and \cite{kirby15}, however, these intrinsic uncertainties do not take into account
other possible sources of systematic effects. The first of these two papers quantified the systematic error to be
2.2 km/s, whereas the second paper reduced it to 1.5 km/s thanks to significant improvements in the standard reduction pipeline.
Since we used the public version of the pipeline, we decided to add in quadrature the additional 2.2 km/s term quoted by \cite{simongeha07}
to the intrinsic uncertainties.

\begin{figure}
\includegraphics[width=\columnwidth]{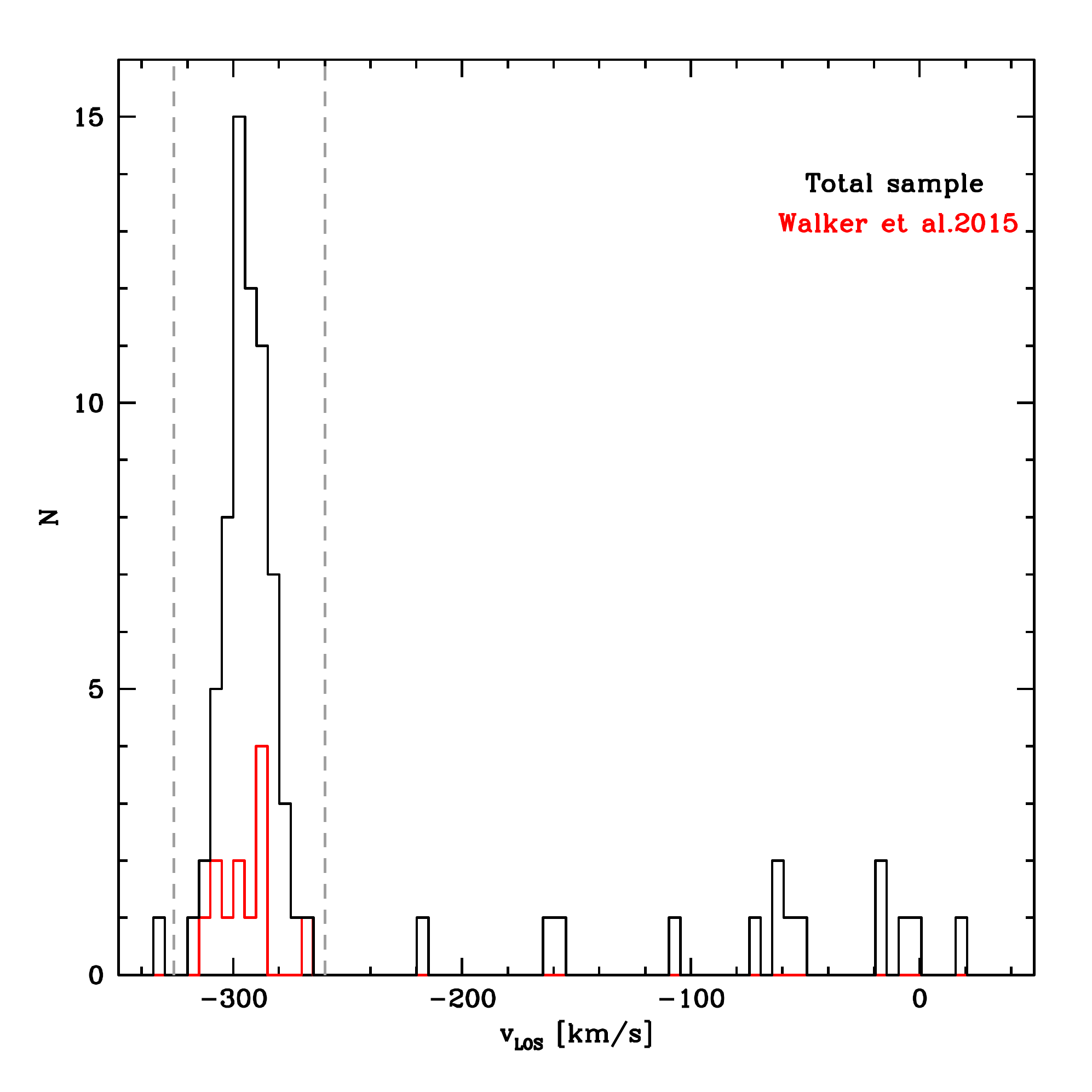}
\caption{\small LOS velocity distribution for the entire sample of 81 stars. The red histogram shows the distribution of targets
with LOS velocities coming from \cite{walker15} only. Grey dashed lines mark the boundaries of the adopted membership criterion.}\label{rv}
\end{figure}

After the application of the offsets and the computation of the total uncertainty $\epsilon^{TOT}_{LOS}$, we computed the v$_{LOS}$ 
for the stars in common between \cite{walker15} and our list using the weighted mean.
Our final v$_{LOS}$ catalogue thus includes $81$ stars (their distribution is the black histogram in Fig.~\ref{rv}), 
of which $51$ had never been targeted before, $18$ are in common with \cite{walker15}, while the remaining $12$ come from \cite{walker15} only (red histogram in Fig. \ref{rv}). 
The peak in the distribution defined by Draco-likely members emerges clearly from the rest and has a mean v$_{LOS}$ of $-293.7\pm1.2$ km/s,
which is in good agreement with the Draco mean velocity of $-292.8\pm0.5$ km/s found using the entire sample of \cite{walker15}.

The list of all the LOS velocities used here is given in Table~\ref{tab1}, and is available in its entirety through the 
Centre de Données astronomiques de Strasbourg (CDS).

\section{Velocity dispersions}

With all the ingredients in hand, we selected the final sample of member stars with 3-D kinematics to be used to
determine the velocity dispersions in the radial, tangential, and LOS direction.
The criteria for a star to belong to such a sample are:
$i$) $-326<$v$_{LOS}<-260$ km/s (see the grey dashed lines in Fig.~\ref{rv}), 
meaning that a member is within $3.5$ times the dispersion around the mean receding velocity of Draco; 
$ii$) that the proper motion of a star differs from the mean proper motion of Draco by 
less than 1 \masyr. This corresponds to $\sim360$ km/s at the distance of Draco and is about a factor of $50$ larger than the typical
errors on the proper motions. Therefore, it is a very weak constraint 
that will not artificially affect the measurement of the velocity dispersion; 
$iii$) \gaia astrometric\_excess\_noise $ <1$ to ensure that a source is a single star (and not an extended
object or unresolved binary);
$iv$) $G>19$ in order to avoid stars that are in the non-linear regime of the HST detector,
where there may be systematic effects on their positional measurements.

We determined  $45$  stars fulfilling these selection criteria and they will be used in the analysis below. 
Table \ref{tab2} lists their positions, magnitudes, proper motions, v$_{LOS}$ , and related errors
for the first $10$ entries and is available in its entirety through CDS. 

We then followed the prescriptions in M18 to determine the velocity dispersion on
the plane of the sky.
First, we transformed the proper motions from the measured equatorial reference to radial and tangential components 
according to the equatorial-polar coordinates relation in \cite{binney08}:
$$\left[ \begin{array}{c} \mu_{R} \\ \mu_{T} \end{array} \right] = \begin{bmatrix} \cos(\phi) & \sin(\phi) \\ -\sin(\phi) & \cos(\phi) \end{bmatrix} \times \left[ \begin{array}{c} \mu_{\alpha}\cos(\delta) \\ \mu_{\delta} \end{array} \right],$$
where $\phi=\arctan(y/x)$, and $x$ and $y$ are the (local Cartesian) gnomonic projected coordinates. Uncertainties are fully propagated,
taking into account the correlation coefficient between the $\alpha$ and $\delta$ positions from {\it Gaia}. The projected velocities in the radial and tangential
direction are, therefore, $v_{R,T} = 4.74 \mu_{R,T} d$, where we assumed $d=76$ kpc to be the
distance to Draco (\citealt{bonanos04}).

We modelled the velocity dispersion for the sample of 45 stars using a multivariate Gaussian and including a
covariance term. The parameters of the Gaussian are the velocity dispersions in the (projected) radial and tangential directions
($\sigma_R, \sigma_T$), their correlation coefficient $\rho_{R,T}$ , and the mean velocities ($v_{0,R}, v_{0,T}$). 
We used the Bayes theorem to derive the posterior distribution for these parameters. We assumed a weak Gaussian prior on the dispersions
(centred on $10$ km/s and with $\sigma=3$ km/s) and on the correlation coefficient
(with mean 0, and dispersion 0.8), and a flat prior for the mean velocities.
The likelihood is a product of Gaussians, and the covariance matrix is the sum of the covariance
matrices associated to the intrinsic kinematics of the population and to the measurement uncertainties (see equation
3 in the Methods Section of M18).

\begin{figure*}
\includegraphics[width=\textwidth]{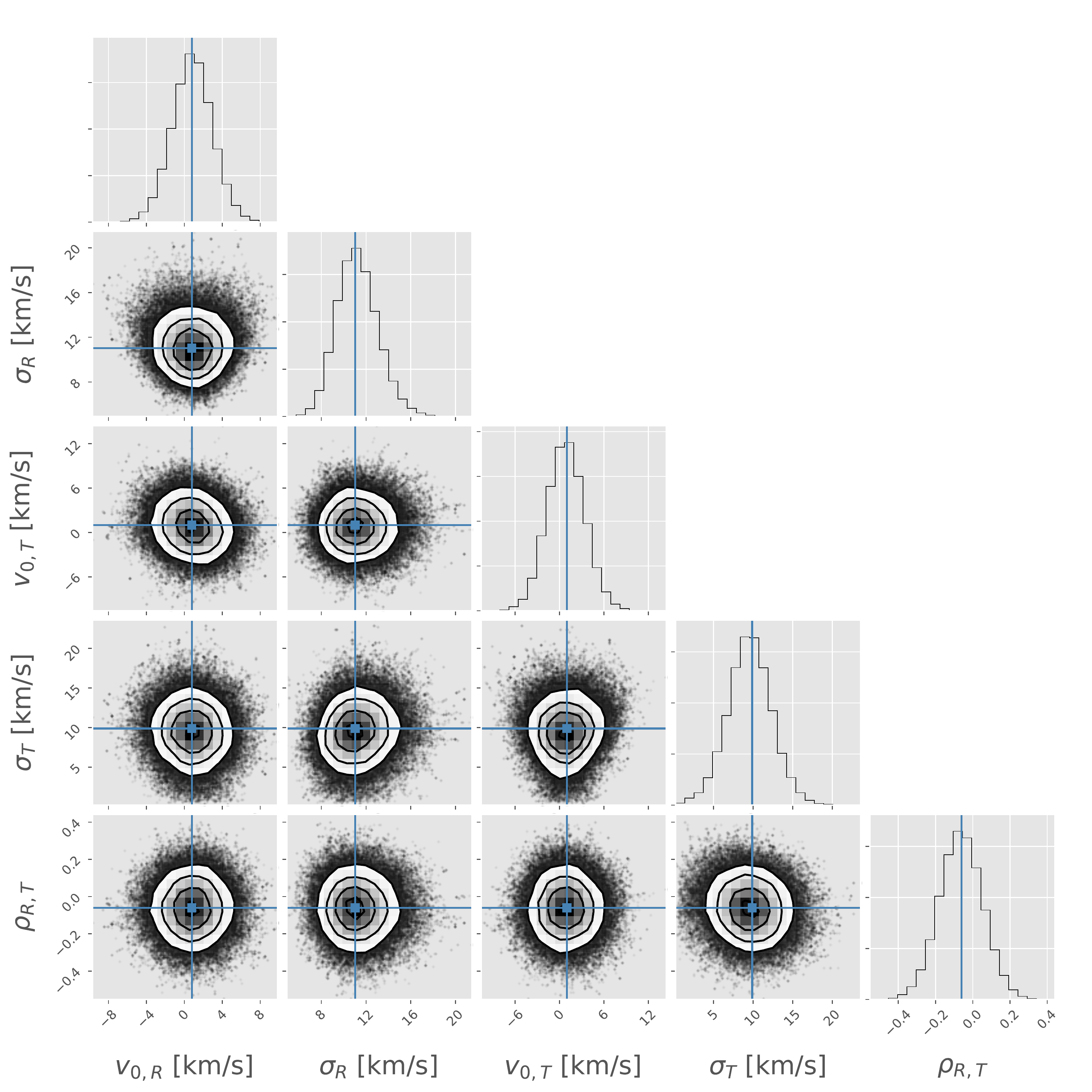}
\caption{\small Corner plot for the modelled parameters. The maximum a posteriori value for each parameter is highlighted in blue.}\label{sigmas}
\end{figure*}

We used the Markov Chain Monte Carlo (MCMC) algorithm from
\cite{foremanmackey13} to estimate the posterior for all the kinematic
parameters and the results are shown in Fig.~\ref{sigmas}\footnote{The code used in this plot has been provided by \citealt{fm16}}.  For our sample, we find $\sigma_{R} =11.0^{+2.1}_{-1.5}$~km/s and
$\sigma_{T}=9.9^{+2.3}_{-3.1}$~km/s, where the quoted errors correspond to
the 16th and 84th percentiles.  In our analysis, we have left the
mean projected velocities $v_{0,R}$ and $v_{0,T}$ as free (nuisance)
parameters, finding these values in good agreement with Draco's estimated
mean motion (\citealt{helmi18}).

To estimate the dispersion in the LOS velocity and its uncertainty, we
applied the maximum-likelihood method described in \cite{walker06},
finding $\sigma_{LOS}=9.0\pm1.1$~km/s.  This value is in good agreement with
the LOS velocity dispersion profiles in \cite{kleyna02, wilkinson04,
  walker15} as per the location of our stars (the mean distance of our
sample from Draco centre is R$_{HST}\simeq5.57\arcmin\simeq120$ pc).
We note that if we remove the subset of $8$ stars with v$_{LOS}$
coming from \cite{walker15} from our final sample of $45$ stars,
$\sigma_{LOS}$ changes by only $0.02$~km/s. This further demonstrates
that the relative calibration between our measurements and theirs
worked adequately.

\section{Dynamical modelling}

We now explore the constraints provided by our measurements on the
internal dynamics of Draco. We focus on the velocity anisotropy
$\beta$ and mass distribution, and, in particular, on the maximum circular
velocity $V_{max}$. In what follows, we\ assume that our
measurements of the velocity dispersions were obtained at the same
location\footnote{Or, alternatively, that they vary slowly within the
  projected radial range probed by the fields in our dataset.}, namely at
R$_{HST}$, which is the average projected distance from the centre of
Draco.

\subsection{A direct measurement of $\beta$}

The velocity anisotropy is defined as
$\beta(r) = 1 -(\sigma_t/\sigma_r)^2$ (\citealt{binney80}),
where $\sigma_t(r)$ and $\sigma_r(r)$ are the intrinsic (3D) velocity
dispersions in the tangential and radial directions, respectively. The
anisotropy $\beta(r)$, the stellar 3D density profile of the stars
$\nu_\star(r)$, and $\sigma_r(r)$ are related to the observed,
projected velocity dispersions at projected distance $R$ as follows
\citep{strigari07}:
\begin{equation}
\sigma_{\rm los}^2(R)=\frac{2}{I_\star(R)} \int_{R}^{\infty} \left ( 1 - \beta \frac{R^{2}}{r^2} \right )
\frac{\nu_{\star} \sigma_{r}^{2} r dr}{\sqrt{r^2-R^2}} \,, \label{eq:LOSdispersion} 
\end{equation}
\begin{equation}
\sigma_R^2(R)=\frac{2}{I_\star(R)} \int_{R}^{\infty} \left ( 1
- \beta+ \beta \frac{R^{2}}{r^2} \right )
\frac{\nu_{\star} \sigma_{r}^{2} r dr}{\sqrt{r^2-R^2}} \,,\label{eq:Rdispersion}
\end{equation}
\begin{equation}
\sigma_T^2(R)=\frac{2}{I_\star(R)} \int_{R}^{\infty} \left ( 1 - \beta \right )
\frac{\nu_{\star} \sigma_{r}^{2} r dr}{\sqrt{r^2-R^2}}  .  \label{eq:phidispersion}
\end{equation}
From these equations, we can derive an estimate
for the anisotropy at a radius $\hat{r}$ using the intermediate
value theorem {\bf(see the full derivation in \citealt{massari18})}: 
\begin{equation}
 \hat{\beta} = \beta(\hat{r}) = 1 - \frac{\sigma^2_T}{\sigma^2_\text{\rm LOS} + \sigma^2_{R} - \sigma^2_{T}}\Bigg\rvert_{{\rm R}_{HST}}.\label{eqbeta}
\end{equation}
with $\hat{r} \in $ [R$_{HST}, r_{tid})$ and where $r_{tid}$ is the
tidal radius of Draco (i.e. the stellar density is zero beyond this
radial distance). This relation is also valid if we assume $\beta$ is
constant (see M18).

\begin{figure}
\includegraphics[width=0.5\textwidth]{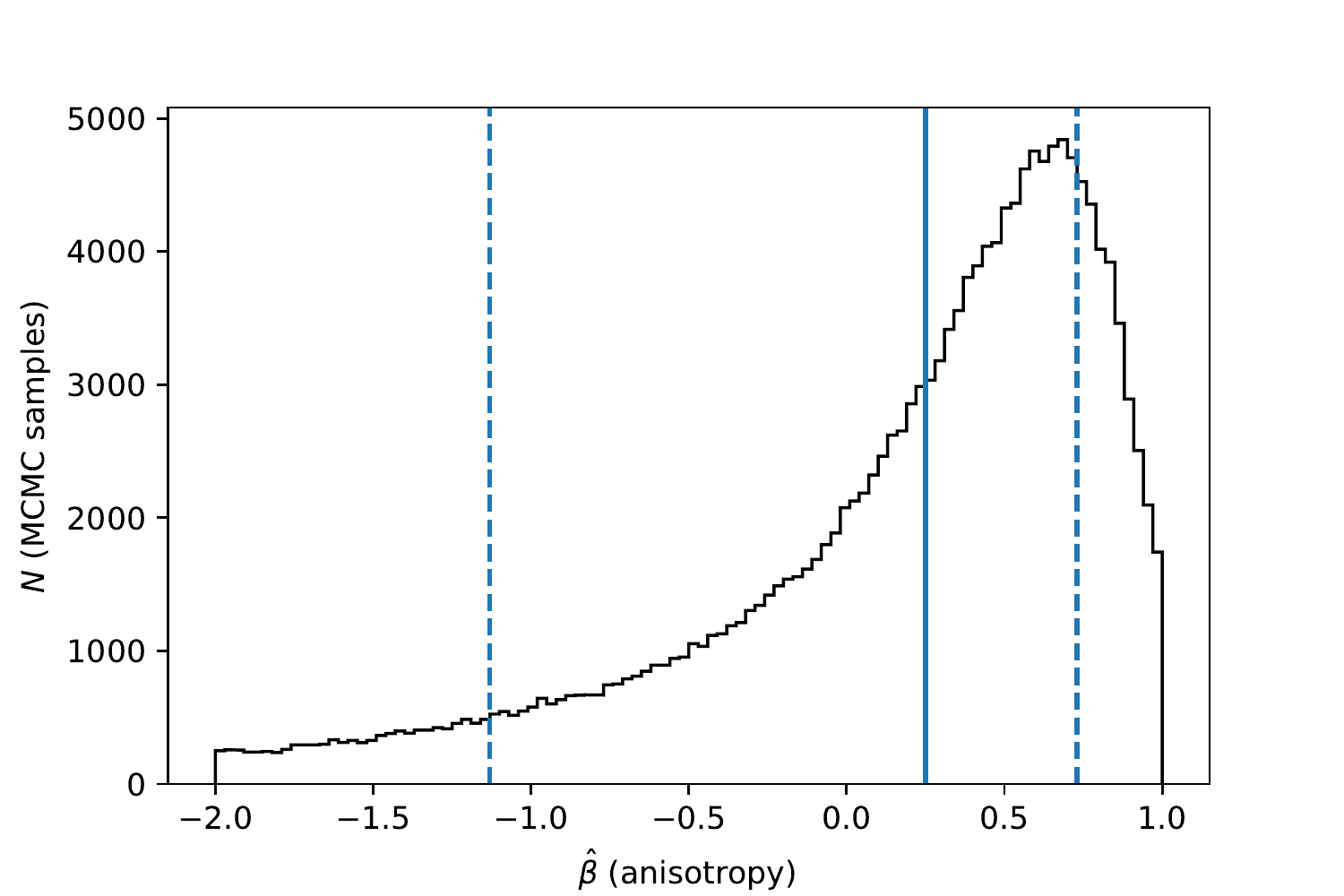}
\caption{\small Posterior distribution for anisotropy
  $\hat{\beta}$ computed using Eq.~\ref{eqbeta}.  The median value is
  highlighted with the vertical solid line, whereas the 16th and 84th
  percentiles are highlighted with blue dashed lines. These have been
  computed using the whole range of possible anisotropies spanned by
  the distribution, even though this figure shows a shorter range for
  visualisation purposes.}\label{beta}
\end{figure}

Fig.~\ref{beta} shows the posterior distribution for $\hat{\beta}$
obtained using our measurements and Eq.~\ref{eqbeta}.  This figure
shows that radial anisotropies are favoured (although the
uncertainties are large), with a median value of
$\hat{\beta}=0.25^{+0.47}_{-1.38}$, and where the lower and upper
limits correspond to the 16th and 84th percentiles of the
distribution, respectively.

\subsection{Joint constraints on $V_{max}$ and $\beta$}

We now use our measurements at R$_{HST}\simeq120$ pc to
simultaneously constrain $\beta$ and $V_{max}$. To this end, we
introduce the Jeans equation for a spherical system
(\citealt{battaglia13}):
\begin{equation}
\label{eq:jeans}
\sigma_{r}^{2} = \frac{GM(r)}{r} \frac{1}{\gamma-2\beta-\alpha},
\end{equation}
where $\gamma=d\log\nu_\star/d\log r$, and
$\alpha=d\log\sigma_{r}^{2}/d\log r$. For Draco, we may assume that
$\alpha<<\gamma$ because its v$_{LOS}$ dispersion profile is known to
be relatively flat \cite[e.g.][]{kleyna02,walker09, walker15}. 

We compute $\sigma_r(r)$ from Eq.~\ref{eq:jeans}, for different mass
models assuming different (constant) values of $\beta$. Replacing
these in Eqs.~\ref{eq:LOSdispersion}, \ref{eq:Rdispersion}, and
\ref{eq:phidispersion} allows us to derive confidence contours in the
characteristic parameters of the models that are consistent with the
measured values of the projected velocity dispersions.

We model the mass of the system as the sum of a dark and a
stellar component. For the stellar component, we adopt a Plummer
profile (\citealt{plummer11}) with a projected half-light radius of
R$_{1/2}=196$~pc (\citealt{walker07}) and stellar mass of
$3.2\times10^5$~M$_\odot$ (\citealt{martin08}). For the dark halo
component, we assume an NFW profile (\citealt{nfw96}), for which
\begin{equation}\label{eqnfw}
 M(r)=4\pi\rho_0r_s^3[f(r/r_s)]=M_sf(r/r_s),
\end{equation}
where $\rho_0$ is a characteristic density, $r_s$ the scale radius, and
$f(x) = \ln(1+x)-x/(1+x)$. The mass $M_s$ and the peak circular velocity
$V_{max}$ are related via 
\begin{equation}
\label{eq:ms}
 M_s=\frac{R_{max}V_{max}^2}{{\rm G} f(2.163)},
\end{equation}
where G is the gravitational constant, and where the peak circular
velocity is at radius $R_{max} =2.163r_s$ for the NFW profile. On the
other hand, the virial mass\footnote{In this paper the virial
radius is defined as the radius at which the average density is 200 times the critical density}  is $M(r_{vir})=M_{vir}=M_sf(c_{vir})$,
where $c_{vir} = r_{vir}/r_s$.

Cosmological simulations have shown that there is a tight relation
between the virial mass of a halo and its concentration $c_{vir} = r_{vir}/r_s$, such that the NFW profiles are effectively a one-parameter family (e.g. \citealt{bullock01}). 
For sub-haloes,  that is, haloes surrounding
satellite galaxies such as Draco, the concept of virial mass is
ill-defined, however.  This is why we prefer to work with the peak circular
velocity $V_{max}$, which has been shown to vary little when a halo
becomes a satellite (\citealt{kravtsov04}). For
sub-haloes, there also is a relation between the mass-related parameter
$V_{max}$ and a concentration parameter $c_V$ defined as \citep{diemand07}
\begin{equation}
\label{eq:cv}
 c_V=2\left(\frac{V_{max}}{H_0 R_{max}}\right)^2,
\end{equation}
where $H_0$ is the Hubble constant, which we assume to be $H_0=70$
km/s/Mpc. Using high-resolution N-body cosmological simulations
\cite{moline17} found that
\begin{equation}\label{vmaxc}
 c_V=c_0[1+b\log(x_{sub})]\left[1+\sum_{i=1}^{3}\left[a_i\log\left(\frac{V_{max}}{\rm10 km/s}\right)\right]^{i}\right],
\end{equation}
where $c_0=35000$, $a_i={-1.38, 0.83, -0.49}$, $b=-2.5$ for a sub-halo
located at the distance of Draco, that is, for d = $76$ kpc and the virial
radius 
of the host halo to be $221.6$ kpc (\citealt{mcmillan17}), for which $x_{sub}=0.34$.

\begin{figure*}[!t]
\includegraphics[width=\textwidth]{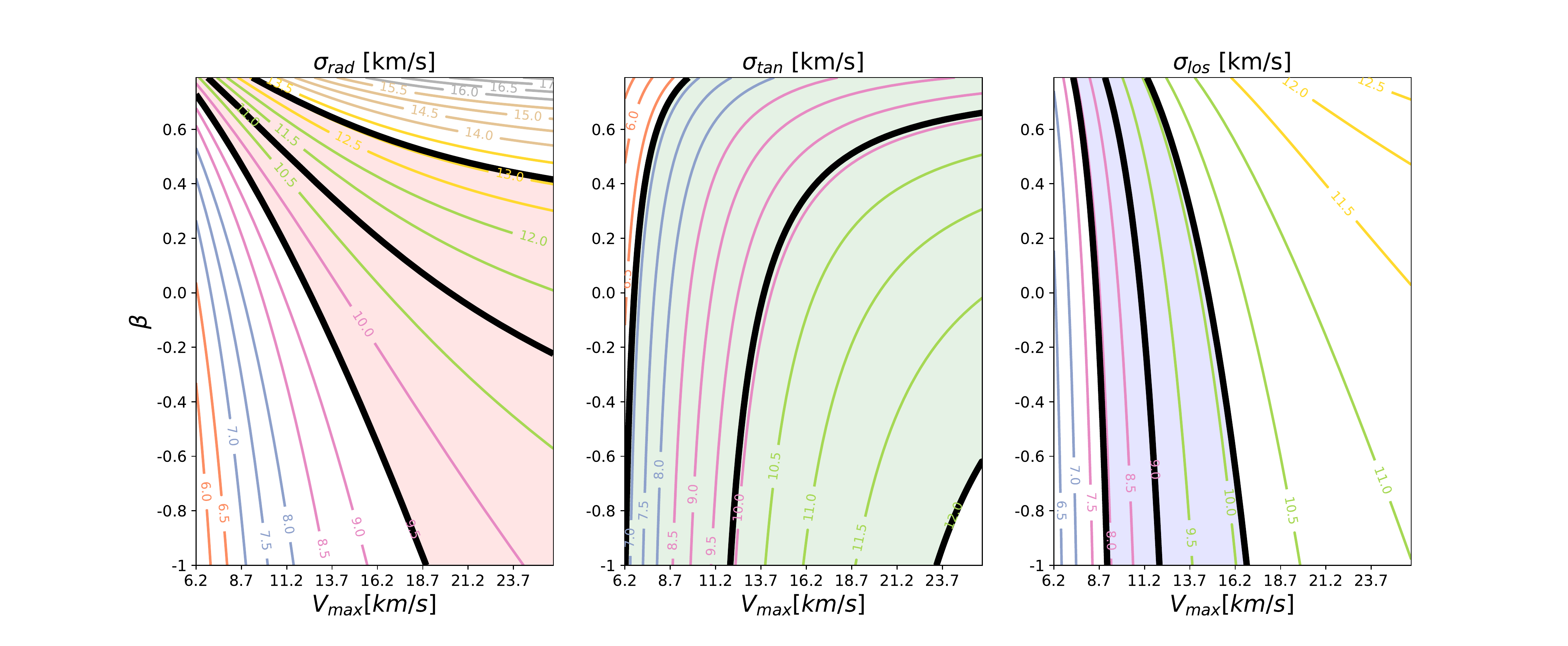}
\caption{\small Velocity dispersion maps in the $V_{max}$-anisotropy
  space. Left, central and right panels show the velocity
  dispersions in the projected radial, tangential and in the LOS
  components, respectively. Black solid lines and coloured-shaded
  areas indicate the range defined by our mean measurements and their
  16th and 84th percentiles, at R$_{HST}$. Coloured lines indicate
  contours of constant dispersion in steps of $0.5$~km/s. }\label{nfw}
\end{figure*}
 
We proceed to sample the space of parameters defined by
($V_{max}$, $\beta$). For each $V_{max}$ , we derive $R_{max}$ (and
hence $r_s$) using Eqs.~\ref{eq:cv} and \ref{vmaxc}, and $M_s$ from
Eq.~\ref{eq:ms}. We then insert Eq.~\ref{eqnfw} (through Eq.~\ref{eq:jeans}) in
Eqs.~\ref{eq:LOSdispersion}--\ref{eq:phidispersion} to compute the
predicted values of $\sigma_{LOS}$, $\sigma_R$ and $\sigma_T$ at
R$_{HST}$ for each pair of input parameters. 

The three panels in Fig.~\ref{nfw} show the results of this
procedure. Coloured lines indicate contours of constant dispersion in
steps of $0.5$ km/s in the space of ($V_{max}$, $\beta$). In each
panel, black solid lines mark the range defined by our mean measurements
and the corresponding 16th and 84th percentiles (also highlighted by the coloured shaded
areas). From Fig.~\ref{nfw} we note that at R$_{HST}$, $\sigma_{LOS}$
gives basically no constraint on $\beta$, whereas $\sigma_R$ is the
most sensitive parameter in this respect. On the other hand,
$\sigma_{LOS}$ provides the strongest information on the $V_{max}$ of
the system, while $\sigma_T$ does so only for tangential anisotropy
and becomes sensitive to $\beta$ for radial anisotropy. 

\begin{figure}
\includegraphics[width=\columnwidth]{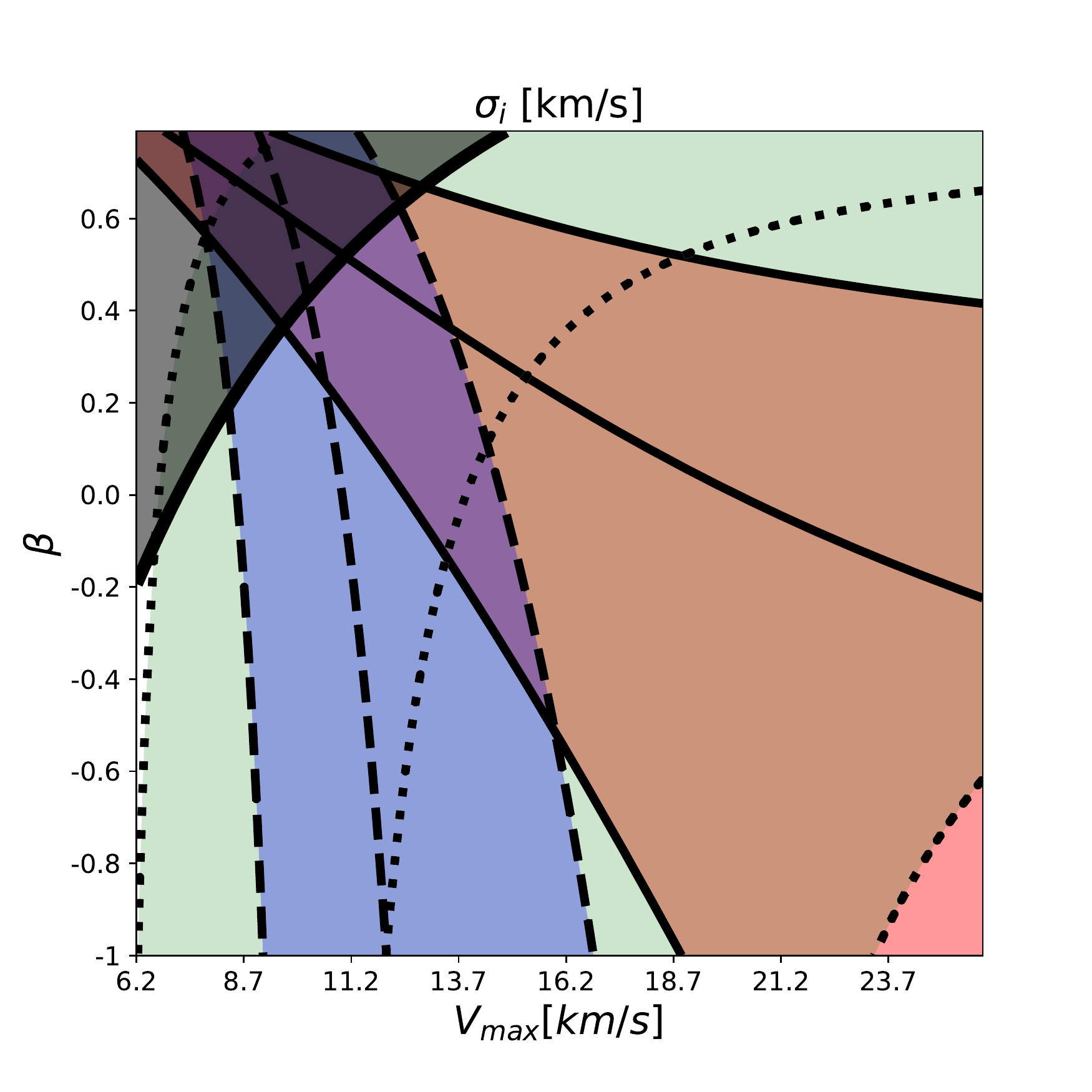}
\caption{\small Three independent constraints coming from the measured velocity dispersions (from Fig.~\ref{nfw}) over-plotted together.
Their intersection (dark purple area) shows the range of allowed values for $V_{max}$ and constant anisotropy $\beta$.}\label{overplot}
\end{figure}

To better highlight the constraints on $V_{max}$ and $\beta$ given by
our measurements, we over-plot the three independent constraints in
Fig.~\ref{overplot}, using different line styles and colour-coding to
demarcate each contribution. The three independent estimates all
intersect in the darkest shaded region. The existence of a common
solution demonstrates that our results are consistent with the NFW
profile predicted by CDM models for the halo of Draco.  

To check whether the entire range of parameters sampled by our solution 
is physically meaningful, we use the escape velocity expected for an NFW halo as a constraint \citep[see Eq.16 from][]{shull14} since $V_{esc}$ is related to $V_{max}$. Assuming stars in Draco follow a Gaussian
velocity distribution, with a characteristic dispersion of 
$\sigma_{3D}=\sqrt{\sigma_{R}^2+\sigma_{T}^2+\sigma_{LOS}^2}$, we require that 
most of our stars in Draco be bound, that is, $3*\sigma_{3D}<V_{esc}$. At r=r$_{HST}$ (where our measurements lie) this constraint results in a large portion of the ($V_{max}, \beta$) plane being excluded. The allowed region is shown as a black-shaded area delimited by 
a thick solid line in Fig.~\ref{overplot}.
When taking this argument into account, we see that
radial anisotropy seems to be preferred, although the allowed region
extends down to $\beta\simeq-0.6$.  This behaviour is fully consistent
with the posterior distribution for $\hat{\beta}$ (Fig.~\ref{beta})
obtained directly from the data, but shows that different regions are
preferred depending on the $V_{max}$ of the halo in which Draco is
embedded. The range of preferred values for $V_{max}$ goes from $10.2$
km/s to $17.0$ km/s.  This agrees within the errors with the previous
estimate from \cite{martinez15} who quote $V_{max}=18.2_{-1.6}^{+3.2}$~km/s, and it is consistent with the lower limits set by the analysis of
\cite{strigari07b}, who find $V_{max}>15$~km/s.

\begin{figure}
\includegraphics[width=\columnwidth]{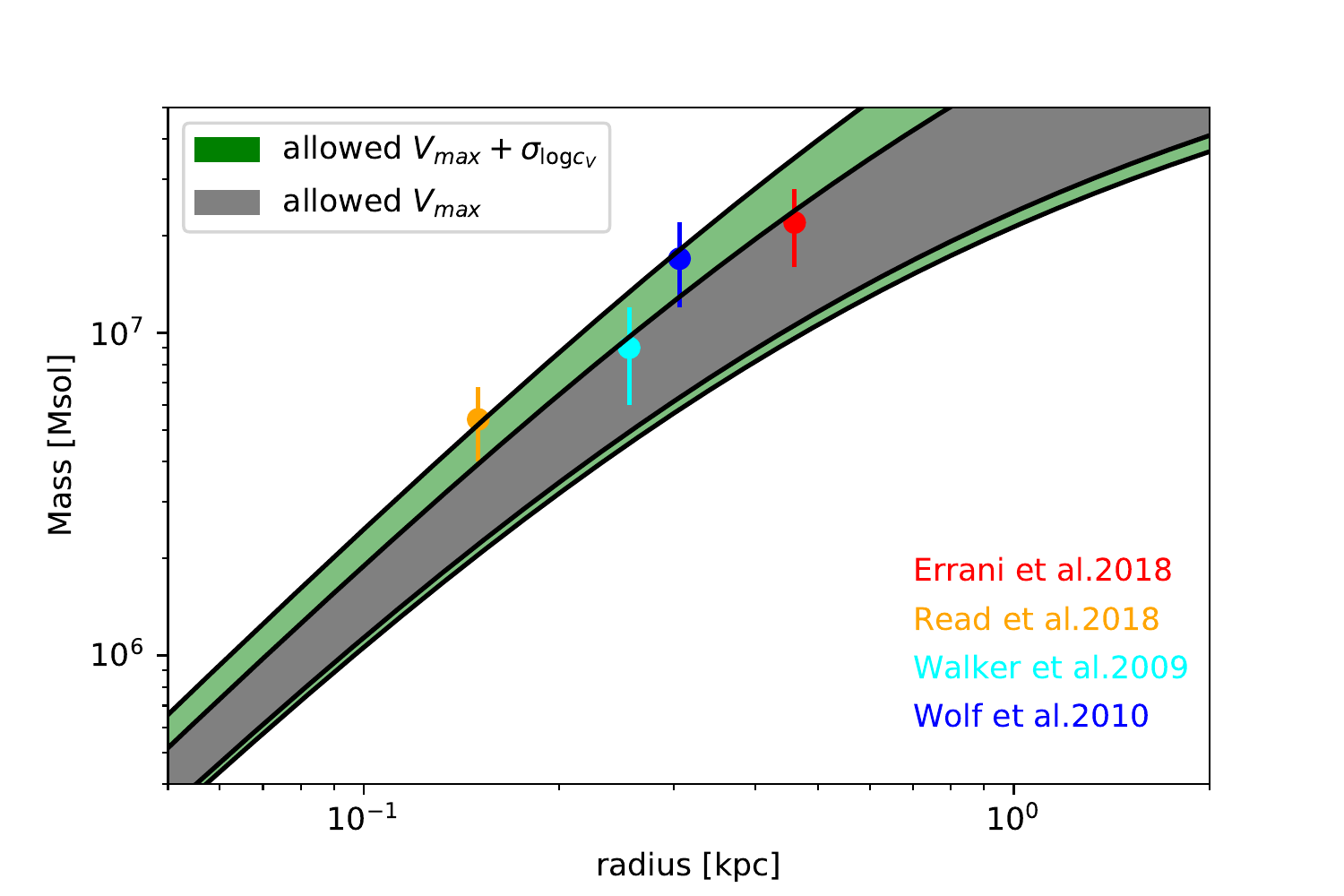}
\caption{\small Comparison between our derived mass profile (grey shaded area)
  and previous mass estimates, colour-coded as indicated by the
  labels. Green-shaded area marks the range of mass allowed by our solution when
  also considering the scatter on the concentration relation.}\label{mass_comp}
\end{figure}

Fig.~\ref{mass_comp} shows the range of dark halo mass profiles for
Draco that best describe our measurements. The grey shaded area has
been derived using the range of $V_{max}$ values allowed by the joint
constraints shown in Fig.~\ref{overplot} (together with the
NFW profile given by Eq.\ref{eqnfw}), while the green-shaded area
corresponds to the range of profiles obtained when also taking into
account the $1-\sigma$ scatter on the concentration relation
(Eq.~\ref{vmaxc}). \cite{moline17} quote a scatter of
$\sigma_{{\rm log}~c_{vir}}=0.11$. Since $c_V$ and $c_{vir}$ are related
through
\begin{equation}\label{eq:cc}
 c_V=200 \left(\frac{c_{vir}}{2.163}\right)^3\frac{f(R_{max}/r_s)}{f(c_{vir})},
\end{equation}
\citep{diemand07}, this implies that 
\begin{equation}
 \sigma_{{\rm log} c_V} \simeq \frac{\delta c_V}{c_V} = \left[3-\frac{c_{vir}^2}{(1+c_{vir})^2}\frac{1}{f(c_{vir})}\right]\frac{\delta c_{vir}}{c_{vir}}, 
\end{equation}
which for the range of $V_{max}$ permitted by our measurements, results in 
$\sigma_{{\rm log} c_V} \simeq 2.54 \sigma_{{\rm log} c_{vir}}$ or $\sigma_{{\rm log}~c_V}=0.28$.

We now compare the results of our mass modelling with published
studies based on the use of line-of-sight velocities (see also
Fig.~\ref{mass_comp}). We first focus
on three robust mass estimators which suffer little from the
mass-anisotropy degeneracy.  \cite{walker09} determine the mass
enclosed within the projected half-light radius R$_{1/2}$, to be
M$_{1/2}=0.9\pm0.3\times10^7$~M$_\odot$. In comparison, we obtain
$0.4\times10^7<$~M$_{1/2}<1.0\times10^7$~M$_\odot$, for the $V_{max}$
range favoured by our measurements. \cite{wolf10} propose a mass
estimator M$_{-3}=3\sigma_{LOS}^2r_{-3}/G$, where $r_{-3}$ is the
radius at which the logarithmic slope of the stellar density profile,
${\rm d}\log\nu_\star/{\rm d}\log r = -3$. For our measured
$\sigma_{LOS}=9.0\pm1.1$~km/s and for $r_{-3}=306$~pc, this yields
M$_{-3}=1.7\pm0.5\times10^7$~M$_\odot$, while the range of
masses allowed by our models is
$0.7\times10^7<$~M$_{-3}<1.3\times10^7$~M$_\odot$.  Note that the upper limit of
our mass range increases to $1.8\times10^7$~M$_\odot$ when considering
the uncertainty on the mass-concentration relation.
Finally, \cite{errani18} provide as an estimator the mass enclosed within
1.8 times R$_{1/2}$, defined as M$_{1.8}=3.5\times1.8$R$_{1/2}\sigma_{LOS}^2/G$.
According to our $\sigma_{LOS}$, this gives M$_{1.8}=2.2\pm0.7\times10^7$~M$_\odot$,
which is well within our range of solutions $1.0\times10^7<$~M$_{1.8}<2.3\times10^7$~M$_\odot$.

Another interesting comparison is made with \cite{read18}, who use the
$\sigma_{LOS}$ profile find a spherically-averaged dark matter density
at $r =150$~pc of $\rho_{DM}=2.4^{+0.5}_{-0.6}\times10^{8}$~M$_\odot$~kpc$^{-3}$, which the authors argue favours the case for a
cusp in Draco.  
In our case, given the allowed $V_{max}$ range and the scatter on the concentration relation,
we find $0.8\times10^{8}<\rho_{DM}<2.3\times10^{8}$~M$_\odot$~kpc$^{-3}$,
which is consistent with their estimate.
It is worthwhile noticing that \cite{read18}
infer a slightly tangential anisotropy and that, indeed, the similarity
with our results is stronger when we consider the high mass side of our
range (i.e. when our $\beta$ is more tangential).  

Therefore, our mass models based on the use of proper motions and
line-of-sight velocities of stars in Draco are in good agreement with
published results from analytical estimators \citep{walker09,wolf10} and with the more sophisticated modelling using the full line-of-sight
velocity dispersion profile \citep[e.g.][]{read18}.  Still, our mass
  estimate tends to be closer to the lower limit of previously
  reported measurements (although well within the uncertainties). This
  could potentially indicate that Draco is more concentrated than
  has been predicted by the median $V_{max}$-concentration relation, which is
  also favoured by the analysis of \cite{read18}.  The latter
  work favoured a cold dark matter cusp for Draco mass profile. Our
  measurements and the consistent density estimate support the same
  conclusion.

\section{Conclusions}\label{concl}

In this paper, we present the first measurement of the velocity
dispersion tensor of the Draco dwarf spheroidal galaxy.  The proper
motions on the plane of the sky were derived combining HST and \gaia
data, following the procedure developed by M18 for the
Sculptor dwarf spheroidal.  We complemented the proper motions with
$51$ new LOS velocity measurements from the DEIMOS spectrograph. After
making a selection based on S/N and likely membership, we constructed a sample
of $45$ stars having 3D velocities of exceptional quality (with typical
errors on the individual 3D velocity $<10$~km/s).  For this sample,
located on average at 120 pc from the centre of Draco, we find
dispersions of $\sigma_{R} =11.0^{+2.1}_{-1.5}$~km/s,
$\sigma_{T}=9.9^{+2.3}_{-3.1}$~km/s, and
$\sigma_{los}=9.0^{+1.1}_{-1.1}$~km/s.  The uncertainties are almost a
factor of two smaller than those in M18 for Sculptor.

These measurements allowed us to derive the posterior distribution of
the orbital anisotropy $\beta$, at a radius $r \gtrsim
$R$_{HST}$. This distribution is extended and peaks at
$\hat{\beta} \sim 0.68$, with a median of
$\hat{\beta}=0.25^{+0.47}_{-1.38}$, where the lower and upper limits
indicate the 16th and 84th percentiles.

We also used these measurements to place simultaneous constraints
on the $V_{max}$ of Draco and its orbital anisotropy (assuming the
latter is constant).  Using the Jeans equations (together with a
requirement that the stars be bound) both for an NFW dark halo and a
Plummer stellar profile, we find $V_{max}$ values in the range
$10.2$ km/s to $17.0$ km/s, which is in good agreement with previous mass
estimates based on LOS velocity measurements only. Although tangential
anisotropy is allowed (up to $\beta \sim -0.6$), the range of allowed
mass models is larger for radial anisotropy.

The fact that a family of solutions for Draco's anisotropy and
$V_{max}$ exist, given our 3D velocity dispersion measurements under
the assumption of an NFW profile, demonstrates consistency with
expectations drawn from cold dark matter models. More detailed dynamical
modelling, together with more precise estimates on the 3D velocity
dispersions, are required to establish firmer
conclusions (see also \citealt{lazar19}). 
Nonetheless, it is particularly encouraging that \gaia, as
it continues its operations, is making proper-motion measurements ever
more accurate. It should soon be possible to determine on more secure
grounds whether the dark haloes of dwarf spheroidal galaxies follow
the predictions of cosmological galaxy formation models.

\begin{table*}
\centering
\caption{List of v$_{LOS}$ measurements for Draco targets. Flag indicates whether the measurement is taken from \citep[][flag$=0$]{walker15}, if it is new (flag$=1$),
or if it comes from the weighted mean between ours and \cite{walker15} measurements (flag$=2$). The entire list is available through the CDS.}\label{tab1}
\begin{tabular}{cccccc}
\hline
$\alpha$ & $\delta$  &   {\it G}   &  v$_{LOS}$   & $\epsilon^{TOT}_{v_{LOS}}$ & flag  \\
  deg    &   deg     &       &  km/s &km/s    &       \\
\hline
    259.8451248562  &  57.9478908265 & 19.380 & -300.7 & 2.5 &  1   \\
    259.8192040232  &  57.9691005474 & 18.930 & -289.8 & 2.3 &  2   \\
    259.8417154779  &  57.9840860321 & 19.197 & -298.4 & 2.6 &  2   \\
    259.8884861598  &  57.9438841701 & 19.318 & -286.1 & 2.5 &  2   \\
    259.8647236300  &  57.9442576339 & 20.178 & -311.0 & 3.8 &  1   \\
    259.8225679085  &  57.9641182530 & 20.191 & -219.3 & 2.9 &  1   \\
    259.9216037743  &  57.9618490671 & 18.346 & -302.5 & 2.6 &  1   \\
    259.8598480665  &  57.9932178145 & 18.599 & -299.4 & 2.8 &  1   \\
    259.9041829235  &  57.9631190310 & 18.954 & -286.6 & 2.5 &  1   \\
    259.9215205028  &  57.9811054827 & 19.194 &   +0.4 & 2.8 &  1   \\ 
\hline
\end{tabular}
\end{table*}

\begin{table*}
\centering
\caption{List of positions, \gaia G-band magnitude, $\mu_{\alpha}\cos(\delta)$, $\mu_{\delta}$, v$_{LOS}$ and related uncertainties for  
  sample of 45 stars with 3D motions used in the dynamical analysis. The entire list is available through the CDS.}\label{tab2}
\begin{tabular}{ccccccccc}
\hline
$\alpha$ & $\delta$  &   {\it G}  &  $\mu_{\alpha}\cos(\delta)$ & $\epsilon_{\mu_{\alpha}\cos(\delta)}$ & $\mu_{\delta}$ & $\epsilon_{\mu_{\delta}}$ &  v$_{LOS}$ & $\epsilon^{TOT}_{v_{LOS}}$  \\
  deg    &   deg     &      &  \masyr    & \masyr         &  \masyr   &  \masyr        & km/s &  km/s   \\
\hline
    259.8417154779 &  57.9840860321 &  19.197 &  -0.027  &  0.034  & -0.154 & 0.028    &  -298.4  &   2.6\\
    259.8568904104 &  57.9515551582 &  19.269 &   0.014  &  0.032  & -0.176 & 0.032    &  -308.5  &   1.4\\
    259.8884861598 &  57.9438841701 &  19.318 &  -0.054  &  0.044  & -0.117 & 0.038    &  -286.1  &   2.5\\
    259.8647236300 &  57.9442576339 &  20.178 &  -0.048  &  0.052  & -0.103 & 0.056    &  -311.0  &   3.8\\
    259.8722406668 &  57.9796046445 &  19.297 &   0.014  &  0.052  & -0.178 & 0.035    &  -309.1  &   2.9\\
    259.8666938634 &  57.9941772891 &  19.392 &  -0.012  &  0.032  & -0.143 & 0.030    &  -290.4  &   3.0\\
    259.9192783256 &  57.9774115703 &  20.069 &  -0.043  &  0.050  & -0.170 & 0.055    &  -298.5  &   2.9\\
    259.8789532265 &  57.9830671561 &  20.150 &  -0.063  &  0.056  & -0.135 & 0.058    &  -315.5  &   2.9\\
    259.9211108657 &  57.9681922763 &  20.201 &  -0.056  &  0.058  & -0.117 & 0.053    &  -302.0  &   3.0\\
    259.8645763982 &  57.9846580785 &  20.204 &  -0.025  &  0.065  & -0.139 & 0.055    &  -286.6  &   4.9\\
\hline
\end{tabular}
\end{table*}

\begin{acknowledgements}

We thank the anonymous referee for comments and suggestions
that improved the quality of our paper.
We are also grateful to Justin Read and James Bullock for useful discussions.
DM and AH acknowledge financial support from a Vici grant from NWO.
LVS is grateful for partial financial support from HST-AR-14583 grant.
LS acknowledges financial support from the Australian Research Council 
(Discovery Project 170100521).
This work has made use of data from the European Space Agency (ESA)
mission \gaia (http://www.cosmos.esa.int/gaia), processed by the
\gaia Data Processing and Analysis Consortium (DPAC, {\tt
http://www.cosmos.esa.int/web/gaia/dpac/consortium}). Funding for
the DPAC has been provided by national institutions, in particular
the institutions participating in the \gaia Multilateral Agreement.
This work is also based on observations made with the NASA/ESA Hubble Space
Telescope, obtained from the Data Archive at the Space Telescope
Science Institute, which is operated by the Association of
Universities for Research in Astronomy, Inc., under NASA contract NAS
5-26555. 
Some of the data presented herein were obtained at the W. M. Keck Observatory, 
which is operated as a scientific partnership among the California Institute of 
Technology, the University of California and the National Aeronautics and Space Administration. 
The Observatory was made possible by the generous financial support of the W. M. Keck Foundation.
The authors wish to recognise and acknowledge the very significant cultural role 
and reverence that the summit of Maunakea has always had within the indigenous 
Hawaiian community.  We are most fortunate to have the opportunity to conduct 
observations from this mountain.

\end{acknowledgements}

\end{document}